\documentclass[aps,prl,twocolumn,groupedaddress,draft,showpacs,showkeys]{revtex4-1}

\usepackage{amssymb}
\usepackage{amsmath}

\begin{document}

% Use the \preprint command to place your local institutional report
% number in the upper righthand corner of the title page in preprint mode.
% Multiple \preprint commands are allowed.
% Use the 'preprintnumbers' class option to override journal defaults
% to display numbers if necessary
%\preprint{}

%Title of paper
\title{Quantum Mechanics without Observers}

% repeat the \author .. \affiliation  etc. as needed
% \email, \thanks, \homepage, \altaffiliation all apply to the current
% author. Explanatory text should go in the []'s, actual e-mail
% address or url should go in the {}'s for \email and \homepage.
% Please use the appropriate macro foreach each type of information

% \affiliation command applies to all authors since the last
% \affiliation command. The \affiliation command should follow the
% other information
% \affiliation can be followed by \email, \homepage, \thanks as well.
\author{William H. Sulis}
\affiliation{McMaster University and The University of Waterloo}
%\email{sulisw@mcmaster.ca}

\date{\today}

\begin{abstract}
The measurement problem and the role of observers have plagued quantum mechanics since its conception. Attempts to resolve these have introduced anthropomorphic or non-realist notions into physics. A shift of perspective based upon process theory and utilizing methods from combinatorial games, interpolation theory and complex systems theory results in a novel realist version of quantum mechanics incorporating quasi-local, nondeterministic hidden variables that are compatible with the no-hidden variable theorems and relativistic invariance, and reproduce the standard results of quantum mechanics to a high degree of accuracy without invoking observers.
\end{abstract}

% insert suggested PACS numbers in braces on next line
% insert suggested keywords - APS authors don't need to do this

%\maketitle must follow title, authors, abstract, \pacs, and \keywords
\maketitle
\section{Preamble}

In spite of nearly a century of intense and profound theoretical and philosophical effort the measurement problem remains one of the outstanding problems in the foundations of quantum mechanics \cite{Wheeler}, raising questions about the nature of reality itself \cite{Norsen,Hemmick}. Its resolution would not only be a major achievement in quantum foundations but might also resolve many outstanding paradoxes such as Hardy's paradox, Schrodinger's cat, delayed choice, and perhaps provide new insight into quantum gravity and the problems associated with the many infinities of quantum field theory.

In this paper I propose a realist, quasi-local, nondeterministic, time directed, hidden variable model without observers which is relativistically invariant, avoids the no-hidden variable constraints, yet is capable of reproducing all of the results of non-relativistic quantum mechanics (NRQM) to a high degree of accuracy. It is at least theoretically testable. In this model, quantum mechanics emerges as the asymptotic limit of an inherently unobservable and intrinsically discrete lower level dynamics. 

Many approaches have been taken to resolve this problem, varying in the degree to which they consider the wave function as complete, incomplete, ontic or epistemic \cite{spekkens}. The wave function may be considered as complete and ontic, in which case additional assumptions are incorporated into the quantum mechanical formalism such as multiple universes, multiple times (stochastic quantization), nonlinear terms (continuous spontaneous localization), noise terms, decoherence. It may be considered complete and epistemic as in the consistent histories approach. It may be considered incomplete and an additional set of \textquoteleft hidden variables\textquoteright $\:$ introduced (Bohmian mechanics) or a lower level dynamics such as cellular automata \cite{wolfram} or causal sets \cite{sorkin} is added. There have also been attempts to reformulate quantum mechanics in different mathematical languages \cite{Coecke} or using non-Kolmogorovian probability theory \cite{Khrennikov}.

Since the seminal writings of Bohr, the measurement problem has generally been conflated with questions about the role of observers, and whether or not a realist model of quantum mechanics is possible. In the absence of an ultimate level of reality, quantum phenomena appear not to possess definite properties without the intervention of an observer. This brings a rather discomfiting aspect of anthropomorphism into physical discourse. To eliminate observers requires the assertion of an underlying reality, but Bell's research appears to show that any such reality must possess rather odd features, such as nonlocality. This paper explores the possibility that the measurement problem and the many paradoxes of  quantum mechanics arise because of the unfortunate choice of functional analysis on Hilbert spaces as the representational setting for its theory. A shift to a different representational system can eliminate these problems while still maintaining a realist metaphysics.
  
The approach taken here is based upon process theory \cite{Whitehead, eastman}, archetypal dynamics and emergence theory \cite{Sulis}, combinatorial game theory \cite{Conway}, forcing \cite{hodges} and interpolation theory \cite{Zayed}. It proposes a Planck or sub-Planck scale dynamics from which quantum phenomena emerge at larger scales. It assumes that states yield bounded measurements and that wave functions are band-limited. Modeled upon Whitehead's \textquoteleft actual occasions\textquoteright , its primitive elements emerge into existence apparently possessing spatio-temporal extension and definite properties, then fade into nonexistence while leaving an informational trace than can be carried forward by subsequent elements. Physical entities are viewed as emergent, informationally coherent patterns of primitive elements. 

The core idea is derived from Whitehead's process theory, and considers all physical phenomena as emergent from an ultimate reality of information laden entities called actual occasions. The basic postulates are that 
\begin{enumerate}
\item everything in reality is generated by process, 
\item everything that we observe is emergent from an ultimate lowest level that itself is inherently unobservable,
\item individual events of reality come into existence in a discrete but non-localized form, and what we observe is a diffuse avatar, which extends over space and time and constitutes a wave function,
\item this wave function is an emergent effect but it is through such wave functions that observable physical phenomena arise. 
\end{enumerate}
Reality possesses two aspects - actual occasions, which are the primitive experiential elements, and processes, which generate the actual occasions. Manifest actual occasions in turn influence which processes are active and interacting.  All physical entities are emergent from these actual occasions. This poses intrinsic limits on the observational abilities of these entities. Individual actual occasions cannot be observed, in general. Process too can be inferred but not directly observed. Processes may be ascribed definite informational parameters which influence the nature of the actual occasions that they generate, which in turn influence the properties ascribed to the resulting emergent physical entities. Processes generate the actual occasions that manifest space and time, but they themselves are to be viewed as having an existence which stands outside of space-time. The quantum nature of reality is a consequence of the discreteness manifesting at the fundamental level while the wave nature of reality is a consequence of the inability to resolve these fundamental events in space and time, thus physical reality acquires emergent wave like aspects. 

Process theory places emphasis upon three important characteristics:

\begin{enumerate}
\item the unfolding of process in the manifestation of actual occasions is strongly determined by the context generated by all participating processes and by the dynamics of their interactions
\item observable aspects of physical reality - continuity, space-time, physical entities, symmetries - are emergent
\item processes interact through multiple forms of linearity
\end{enumerate}

It is commonly asserted that there are three fundamental differences between quantum phenomena and classical phenomena. Quantum mechanics demonstrates:
\begin{enumerate}
\item the quantization of exchange in interactions
\item the existence of Non-Kolmogorovian probability
\item the existence of non-local influences
\end{enumerate}

A major goal of this paper is to show that these features may be true of classical systems as well, and their putative absence is actually a reflection of the mathematical languages that have been used to describe the classical and quantum worlds rather than being properties of those worlds in themselves. There are alternative formulations of dynamics based on iterated function systems and combinatorial games that can describe the classical world and yet manifest these so-called quantum aspects. An argument is given for an expansion of the language used to describe fundamental processes beyond that of functions and functions spaces to include iterated function systems and combinatorial games. In so doing, many of the apparent paradoxes of quantum mechanics disappear without the need to resort to anthropomorphism, solipsism, and bizarre metaphysical constructs, and more importantly, without the need to abandon realism.

\section{Introduction}
\section{The Measurement Problem}

In physics a physical system, at each point along its world line, is assumed to exist in a \textit{state}, and  this state determines all past and future states (at least in the absence of interactions with other systems). In classical physics it is assumed that this state can be described, usually in the form of a set of parameters linked to physical constructs such as position, linear momentum, angular momentum, mass, energy, etc. and that a knowledge of the values of these parameters for a given state suffices to predict these values for past and future states. Such a description of a state is said to be \textit{complete}, since no additional information is required to enable these predictions to be carried out. In classical physics these parameters are assumed to be intrinsic ontological characteristics of the system itself, independent of any other system. It is further assumed that there exist particular systems termed \textit{measurement apparatus} that are capable of interacting with the system in such a way as to manifest the values of these parameters without inducing any change in the system. Moreover, it is assumed that all such parameters can, in principle, be simultaneously measured so that it is possible, at least in principle, to actually obtain this complete description of the state of a system. In classical physics it is possible to assert that a physical system \textit{is}, and that it \textit{has} these properties as defined by these parameter values.

The development of quantum mechanics raised serious questions about the validity of these ideas. The discovery by Heisenberg of the uncertainty relations appeared to place fundamental limits on the degree of accuracy with which measurements could be simultaneously carried out. It appeared to be impossible to determine a complete set of parameter values attributable to a given state. 
Bohr suggested that measurements could not in principle be separated from the conditions under which they were obtained. He wrote \cite[pgs. 73,90]{Bohr}

\begin{quote} … every atomic phenomenon is closed in the sense that its observation is based on registrations obtained by means of suitable amplification devices with irreversible functioning such as, for example, permanent marks on the photographic plate, caused by the penetrations of electrons into the emulsion...… the quantum mechanical formalism permits well-defined applications referring only to such closed phenomena and must be considered a rational generalization of classical physics.\end{quote}

Fundamental entities appeared to manifest contradictory properties, on the one hand appearing to be distributed in space-time while on the other hand manifesting effects that appeared to be localized - the so-called wave-particle duality. Properties appeared to occur discretely in some cases, continuously in another, such as energy levels in bound and free particles respectively. It appeared possible to create identical copies of systems that on measurement yielded different values for some parameters, such as occurs in superposition states.

All of these features depart profoundly from the classical conception. In quantum mechanics it appeared that the best one could do was to ascribe a probability distribution for the results of any measurement procedure
applied to a quantum system. The notion of a state changed from being a complete description of the parameter values to being a complete specification of the probability distributions associated with measurements of these values.
While classical physics could assert that at the lowest level of reality there existed definite \textit{somethings} that possessed definite properties whose values were determined by measurements, quantum mechanics could only assert that certain experimental arrangements yielded stable probability distributions of measurements but the status of any underlying ultimate reality remained in doubt. A probability distribution is not a thing in itself but rather is a description of frequencies of occurrences of things. But it is not clear in quantum mechanics who owns these things. Are they actual properties of the physical system being measured? If so then ones encounters a decidedly non-realist view of reality since quantum systems can manifest multiple values of the same property while having indeterminate values of other properties. Or are these so-called properties merely outcomes of certain formalized types of interactions so that they cannot be attributed to either the quantum system or the measurement apparatus but only to the context of the interaction, as Bohr has suggested? But if so, then in what sense does the quantum system possess reality? What is its ontological status when it is not in a measurement interaction? How do two quantum systems interact and what determines such interactions if not these properties and if it is these properties then how do the quantum systems determine what they are in the absence of a measurement interaction?

Problems associated with the ontological and epistemological status of quantum concepts have plagued the theory since its inception and continue to this day. The Copenhagen interpretation of Bohr emphasized the contextuality of measurement. Each measurement was to be understood in the context of a quantum system interacting with a \emph{classical} measurement apparatus, and served to link certain aspects of quantum systems to classical measurement constructs. Unfortunately the pernicious tendency to ignore the emergent and contextual nature of classical properties and to view such properties as intrinsic to entities led to the persistent notion that quantum properties do not exist unless there is an observer and an observation to manifest them.

It is this feature that led to Wheeler's famous dictum that \textquotedblleft No elementary phenomenon is a phenomenon until it is a registered (observed) phenomenon\textquotedblright \cite[pg. 184]{Wheeler}.   

Manifestly though, reality exists without observers. The universe existed long before there were living entities to observe it, and why should living entities possess such a privileged status when they are physical entities like all other physical entities?

There are two very specific problems associated with the notion of measurement and that constitute the so-called measurement problem. The first is the problem of \textit{wave function collapse}. 

Non-relativistic quantum mechanics posits that the state of a physical system is describable by a wave function $\Psi$ which is governed by the Schrodinger equation
 
\begin{displaymath}
i\hslash \frac{\partial\Psi}{\partial t} = H\Psi
\end{displaymath}

\noindent where $H$ is an operator version of the classical Hamiltonian function for the system. $H$ is a linear operator, which is significant because it means that if $\Psi_{1}$ and $\Psi_{2}$ are both solutions of the Schrodinger equation then so is $w_{1}\Psi_{1}+w_{2}\Psi_{2}$ for arbitrary complex weights $w_{1},w_{2}$. 

Measurement is described entirely differently. To each measurement situation there corresponds an operator, $A$ and the possible results of a measurement are given by the eigenvalues of this operator, i.e. those values which satisfy the equation

\begin{displaymath}
A\Psi = \lambda\Psi
\end{displaymath}

Suppose that $\Psi_{1}$ is a solution corresponding to eigenvalue $\lambda_{1}$ and $\Psi_{2}$ is a second solution corresponding to eigenvalue $\lambda_{2}$. Then the sum is a solution, but what value is ascribable to a measurement of such a system? Applying $A$ to the sum yields 

\begin{displaymath}
A(\Psi_{1}+\Psi_{2})=A\Psi_{1} + A\Psi_{2} = \lambda_{1}\Psi_{1}+\lambda_{2}\Psi_{2} \neq \lambda(\Psi_{1}+\Psi_{2})
\end{displaymath}

The sum is \textit{not} an eigenvector of $A$ so it would appear that no measurement can be ascribed to the superposition. In fact, if such a superposition is measured it does yield definite measured values, $\lambda_{1}$ and $\lambda_{2}$ but these appear randomly with frequencies proportional to $<\Psi^{*}_{1}|\Psi_{1}>$ and $<\Psi^{*}_{2}|\Psi_{2}>$ respectively. 

The time evolution of a quantum system is determined by a (linear) unitary operator, so that the time evolution of a superposition under Schr\" odinger evolution always remains a superposition - so how can it ever yield a definite measurement? This was a problem recognized by von Neumann \cite{vonNeumann} and later discussed by London and Bauer\cite{London}. The long held conventional solution is that somehow as a result of a measurement the wave function undergoes a discontinuous transition from the superposition to one of the eigenstates that comprise the superposition. This transition occurs probabilistically with transition probabilities given by the norm of the respective eigenfunction.

There is nothing, however, in the Schr\"odinger form of quantum mechanics that allows for such a \textquoteleft collapse\textquoteright$\:$ of the wave function. It appears neither in the Schr\"odinger formulation nor in the measurement postulate. It must be added separately. Thus this is an ad hoc assumption that to date lacks any physical explanation despite many attempts\cite{Ghirardi}\cite{Everett}\cite{Griffiths} to give it one, including introducing decay factors into the wave function, nonlinearities into the Schrodinger equation, positing the existence of quantum splitting and multiple universes, or the presence of superselection rules that limit solutions. 

The notion of measurement per se is a classical construct, founded upon the behaviour of classical systems in interaction with measurement apparatus. There is no reason a priori that this classical measurement notion should apply to quantum systems. Moreover, unless one is a dogmatic reductionist, there is no reason a priori to believe that the laws of quantum mechanics should apply to classical systems, particularly given the emergent nature of classical systems. These are two realms which, when separated by sufficiently large spatial and temporal scales, manifest distinct features and possess quite distinct mathematical descriptions. Their \emph{effective} theories are quite distinct. Indeed, the archetypal dynamical perspective described in a later section begins with the understanding that the classical and quantum realms constitute distinct categories of reality and require distinct frames of reference for their interpretation and to formulate interactions within them. The quantum mechanical notion of measurement is an attempt to bridge the quantum and classical realms. As Bohr repeatedly emphasized, the idea of measurement is a classical construct and the result of a measurement is a classical object, to whit, a mark, that can be apprehended by a human observer. 
The first problem thus arises due to an attempt to apply classical concepts in a top-down fashion to the quantum realm. The second problem arises from an application of the reductionist paradigm and the attempt to apply quantum concepts to the classical realm. This second aspect arises from the attempt to treat the measurement apparatus as a quantum mechanical object. As London and Bauer demonstrated, the assumption that the measurement apparatus has a description as a quantum mechanical entity simply results in a more complicated superposition of states, now including both the system and the measurement apparatus. The entire coupled system now evolves according to a more complicated wave equation, which, by virtue of its linearity, requires an observer to collapse it. Incorporating this new observer into the situation leads to an infinite regress, a reductionist nightmare.

Shimony showed \cite[pg. 34]{Shimony} that even with approximate measurements, the quantum state of a system will not undergo a transition to an eigenstate under the usual Schrodinger dynamics.

The apparent necessity for some nonphysical process to intervene in order to collapse the wave function has led to several schools of thought. One school proposes that reality requires the active participation of an observer to make it manifest. This introduces an element of psychology (and sometimes mysticism) into physics. A second prominent school of thought suggests that the wave function is a mathematical convenience which enables the calculation of various correlations and functions but which has no counterpart in reality. Some view the wave function as reflecting our ignorance of knowledge of a system, but who's ignorance and why is every observer's ignorance exactly the same? There is an alternative view, however, that treats the wave function as if it were an actual physical wave, with a physical phase \cite{Mann}. Indeed a recent paper \cite{Barrett} has proposed a theorem which appears to demonstrate that the wave function must be real in order to obtain the standard quantum mechanical results. But in that case what does it represent? A distribution of probability? Probability of what? How can a probability be physical? A distribution of mass? Of charge? How does one reconcile that with observed point-like nature of fundamental particles? If the wave function is real, then how to account for the observed quantum nature of quantum interactions? Energy is exchanged in quantum packets but how is this to occur if there is a physically real wave function distributed over space and time? How is it that only one part of the wave contributes to the transfer of energy when both systems have waves that are equally distributed? Why, if the wave is in a superposition of individual eigenfunctions, does the result of a measurement leave the wave in a single eigenfunction? How does such a transformation occur? How can a particle have multiple energies yet transfer only a specific energy?

A less prominent school of thought is that of realism, which holds that at the level of ultimate reality there are actual entities and that quantum mechanics is not the final theory, but needs to be supplemented by a deeper theory of so-called \emph{hidden variables}. Debates about such hidden variables go back to Einstein and Bohr and continue today in questions concerning whether or not hidden variable models are even capable of reproducing the results of quantum mechanics.
 
\section{Realism, Hidden Variables and Nonlocality}

The heart of the problem lies in the question whether the elements of NRQM describe physical reality or are mathematical conveniences that simply enable calculation. This debate has raged since the time of Einstein and Bohr and was thought to be resolved in the famous hidden variable theorems of Bell \cite{Bell} and their subsequent experimental verification beginning with the experiments of Aspect \cite{Aspect}. 
Realist models of quantum mechanics have been around since De Broglie first proposed the idea of a \textquoteleft pilot wave\textquoteright, later developed by Bohm into a fully realized ontological model \cite{Bohm}. The basic idea is to rewrite the wave function as a product, $\Psi = R\exp (iS\hslash)$ where R and S are subject to the differential equations

\begin{displaymath}
\frac{\partial S}{\partial t} + \frac{(\nabla S)^{2}}{2m} + V -\frac{\hslash^{2}}{2m}\frac{\nabla^{2}R}{R} =0 
\end{displaymath}
which can be understood as a Hamilton-Jacobi equation for a particle with momentum $\mathbf{p} = \nabla S$ moving normal to a wave front $S = \text{constant}$ under an additional quantum potential $Q = -\frac{\hslash^{2}}{2m}\frac{\nabla^{2}R}{R}$ while
\begin{displaymath} 
 \frac{\partial R^{2}}{\partial t} + \nabla \cdot \left( R^{2}\frac{\nabla S}{m} \right) = 0
\end{displaymath}
describes the conservation of probability for an ensemble of such particles having a probability density $P=R^{2}$.
 
Bohm interprets these equations as representing the motion of a definite particle coupled to a quantum field given by $\Psi$ (which satisfies the Schrodinger equation). The equation of motion of the particle is given by

\begin{displaymath}
m\frac{d\mathbf{v}}{dt} = -\nabla (V) - \nabla (Q)
\end{displaymath}

The quantum potential does not depend upon the amplitude of the wave in the sense that multiplying the wave function by a constant does not change the value of the quantum potential. It depends only upon the \textit{form} of the wave. A similar situation pertains in signal theory where the information content of a signal depends upon its form but not upon its intensity. Bohm and Hiley make the important point that the quantum potential acts upon the particle not as a form of energy but rather as a form of \emph{information}. Such information is not to be understood in the Shannon sense as a reduction of uncertainty but rather in the semantic sense as providing knowledge about the physical environment within which the particle is moving. This usage of information is uncommon in physics and engineering but virtually universal in most other fields of human endeavor.

Bohm's theory is generally described as a hidden variable model although as Bohm himself points out there are actually no hidden variables in the model. Rather the most important aspect of Bohm's model is that it provides a realist ontological interpretation of quantum mechanics in its assertion that actual particles exist whose motion is guided by the addition of the quantum potential which imparts a stochastic character to their motion  as a result of nonlinear effects.

This quantum potential is decidedly nonlocal in its effect which appears to occur instantaneously. This apparent violation of relativistic constraints has limited acceptance of Bohm's approach but to be fair nothing is being stated as to the ontological character of this quantum potential. While relativity limits the speed at which any signal may propagate there is no suggestion that the information provided by the quantum potential propagates as a signal.

The issue of nonlocality is often raised in association with hidden variable models of quantum mechanics. This will be discussed below but hidden variables are not necessary for nonlocality to arise as an issue in quantum mechanics. It appears even in non-relativistic quantum mechanics in the case of entanglement. For simplicity consider the case of photon pair production through parametric down conversion. Photon polarization gives rise to two possible states $|0>$ and $|1>$. Pair production results in an entangled state of the form $\Psi = \frac{1}{\sqrt{2}}|0>|0> + \frac{1}{\sqrt{2}}|1>|1>$. Measurement of the polarization of one photon automatically guarantees the state of polarization of the second photon, no matter how far the two photons are separated spatially. Experiments have demonstrated that the speed with which this transfer of information occurs is well in excess of the speed of light \cite{Zeilinger}. Shimony has called this a non controllable nonlocality, meaning that it cannot be used to send a signal between two space-like separated observers. This stands in distinction to a controllable nonlocality, which would violate the relativistic constraint. Nonlocality appears to be a feature of the quantum realm regardless of whether or not hidden variables might appear at some lower level.

Bell \cite{Bell} considered the possibility of there being a set of hidden variables from which quantum mechanical correlations might be derived. In his formulation he did not require that these hidden variables be deterministic. In fact he explicitly stated \textquotedblleft It is a matter of indifference in the following whether $\lambda$ denotes a single variable or a set, or even a set of functions, and whether the variables are discrete or continuous.\textquotedblright \cite[pg. 15]{Bell}. The issue of whether or not these hidden variables are deterministic is not germane to his argument. In fact in proving his result it is necessary to assume that these variables are distributed with some probability distribution $\rho$. Whether this distribution arises due to ignorance on the part of the observer or some inherent stochastic dynamics is irrelevant. 

There is a subtlety here which has been discussed in detail by Bunge \cite{Bunge}. There is a notion of determination, which refers to how physical entities acquire their properties, and the mechanisms of such determination, which may be deterministic, non-deterministic or stochastic. The realist view is an assertion that the properties of physical entities are determined in advance, regardless of the mechanism. Hidden variables provide a vehicle to explain how such determination arises. The non-realist view is an assertion that these properties are not determined in advance but arise solely out of the interaction with some observer. In physics, the notion of determination is often conflated with the notion of a deterministic mechanism but that is an error according to Bunge. The requirement for deterministic mechanisms is more ideological than ontological since the discovery of deterministic chaos has blurred the boundary between deterministic and stochastic mechanisms. The real issue is whether or not at the lowest levels of reality there exist actual entities which can be ascribed values given by these hidden variables and which subsequently determine physical entities.

In his argument Bell considered a pair of spin entangled particles $a,b$. Assume that spin measurements are made in the directions $\mathbf{a},\mathbf{b}$ and that along these directions the spins take the values $\pm 1$. Let $E(\mathbf{a},\mathbf{b})$ be the expectation value of the product of the measured values for $a$ in the direction $\mathbf{a}$ and for $b$ in the direction $\mathbf{b}$. Let $\mathbf{c}$ be a third direction. Bell assumes first of all that, given some probability measure on the set of hidden variables, the expectation value of the product is calculated as 

\begin{displaymath}
E(\mathbf{a},\mathbf{b}) = \int_{\Lambda} v(a,\mathbf{a})v(b,\mathbf{b})d\lambda
\end{displaymath}

\noindent where $v(a,\mathbf{a})$ gives the value measured in system $a$ at angle $\mathbf{a}$.

Bell showed that, if an additional assumption of locality is made (meaning that the expectation values of one system must be independent of both the angles measured and the measurement received in the other system), then the following inequality holds

\begin{displaymath}
1 + E(\mathbf{b},\mathbf{c}) \geqq |E(\mathbf{a},\mathbf{b})-E(\mathbf{a},\mathbf{c})|
\end{displaymath}

Nonlocality plays a central role in providing constraints upon the nature of hidden variables that might be involved in determining quantum phenomena. Since Bell's original work, two specific forms of nonlocality have been identified, parameter independence and outcome independence. Parameter independence means that the behaviour of one system does not depend upon the particular choice of property to be measured on another system. Outcome independence means that the behaviour of one system does not depend upon the particular outcome of a measurement of another system. As Shimony points out \cite[pg. 90]{Shimony}, a violation of parameter independence would permit a form of controllable nonlocality which could lead to a violation of relativistic constraints. A violation of outcome independence can at most lead merely to a form of non controllable nonlocality, and indeed the phenomenon of entanglement provides a prima facie case of violation of outcome independence.

The existence of entanglement demonstrates that any model of quantum phenomena will of necessity need to manifest some form of nonlocality. Bell's theorem provides additional evidence for nonlocality by demonstrating that any model of quantum mechanics involving hidden variables that possesses both forms of locality will of necessity lead to predictions that will violate the inequalities. These inequalities  generally involve systems that are separated spatially. Bell's work involved space-like separated systems and was extended by Leggett and Garg \cite{Leggett} to include systems that are separated temporally. While the phenomenon of entanglement demonstrates that nonlocality must be a feature at the observable level, the Bell and Leggett-Garg results show that nonlocality must be a feature of any model at any level. 

There is, however, a subtle assumption in the formulation of Bell's Theorem which has never been questioned until recent times. This assumption is that the probability theory to be associated with any classical process must be Kolmogorovian in nature, so that the formula for the calculation of the individual correlation functions must be Kolmogorovian in form. It is well known that the corresponding formula in the quantum mechanical case is non-Kolmogorovian, as it contains an interference term. The assumption that a classical system must necessarily follow a Kolmogorovian probability structure does not seem to be questioned  in the literature on Bell's theorem. Accepting that assumption then forces any hidden variable model to possess an inherent non-locality. But if this assumption is false, as has been conjectured by Palmer \cite{Palmer} and Khrennikov \cite{Khrennikov} as discussed below, then the conclusion that a hidden variable model \textit{must} be non-local no longer holds. The following section discusses the idea of non-Kolmogorovian probability and demonstrates that it is not necessarily a perquisite of quantum mechanics and can apply to classical systems as well, especially those generated by iterated function systems and games. This will suggest that a change in the model used to express the underlying dynamics to include iterated function systems or game based dynamics might open the door to realist local or quasi-local models.

\section{Non-Kolmogorovian Probability}

The concept of probability is fundamental to the interpretation of quantum mechanics. It is also the source of much of the conceptual confusion and paradoxes that confound quantum mechanics. Hidden variable theorems based upon Bell type inequalities involving relations between various correlation functions depend intimately upon the structure of the probability theory within which these functions are defined and constructed. Indeed that is the very point used by Palmer \cite{Palmer} in his demonstration of a hidden variable model of quantum mechanics using iterated function systems. In that model he showed that it was impossible to construct the required three state correlation functions on account of chaotic effects. As a result the Bell inequalities could be defeated. 

Most people are used to probability theory based upon the axioms of Kolmogorov. The situation in probability today is similar to that in geometry in the 19th century. For two thousand years, Euclidean geometry had held sway as the one true model of geometry. This conception eventually gave way to the realization that there were actually many different types of geometry, just as there turned out to be many different forms of logic and of set theory. It turns out that there are many different forms of probability theory as well. Probabilities associated to classical events are generally held to be modeled exclusively by Kolmogorov type probability theory. That assumption forms a necessary part of the creation of the various Bell inequalities. That this assumption turns out not to be true forms the central thesis of this section.

Probability theory began in the 17th century in the correspondence between Pascal and Fermat on games of chance. There the goal was to aid the gambler in making decisions that would enable them to maximize their profit from playing these games. Probability was linked directly to the idea of frequency. The probability of an outcome was the limiting value of the fraction of times that the event occurred during the play of the game as the number of plays was extended to infinite time. Any individual play of a game resulted in fractions that varied from this number but with repeated play the average of these fractions would tend to the limiting value. This is a consequence of Gauss\textquoteright s celebrated law of large numbers which shows that in repeated measurements of this value the distribution of individual measurements of the value follows a normal distribution. The beautiful mathematical properties of the normal distribution led to its widespread misapplication for more than a century even though natural phenomena manifest events that follow a diversity of probability distributions. This has had serious consequences in fields such as psychology, medicine and economics \cite{West} 

Probability theory was placed on firm mathematical ground in the early 20th century by Kolmogorov, who formulated a set of formal axioms based on set theory and analysis, and subsequently elaborated by Carath\'eodory with his work on measure theory. Crucial to Kolmogorov's theory are three axioms.

\begin{enumerate}
\item Let $\Omega$ be a set and $E \subset \cal{P}$$(\Omega)$. A probability measure $P$ is a map from $E$ to $[0,1]$ such that $P(\Omega)=1$ and $P(A_{1} \cup \ldots \cup A_{n} \cup \ldots ) = P(A_{1}) + \ldots + P(A_{n}) + \ldots$ for disjoint sets $A_{i}$
\item The conditional probability of $A$ given $B$ is defined as $P(A|B) = P(A\cap B)/P(B)$
\item Events $A,B$ are said to be independent if $P(A \cap B) = P(A)P(B)$.
\end{enumerate}

The definition of conditional probability above yields the following formula of total probability: For any partition $\{A_{i}\}$ of $E$, and any $B \in E$,, $P(B) = P(A_{1})P(B|A_{1}) + \ldots + P(A_{n})P(B|A_{n}) + \ldots$. 
Quantum mechanics changed this profoundly. Feynman \cite{Feynman} asserts that \begin{quote} The concept of probability is not altered in quantum mechanics. When we say that the probability of a certain outcome of an experiment is \textit{p}, we mean the conventional thing, i.e., that if the experiment is repeated many times, one expects that the fraction of those which give the outcome in question is roughly \textit{p}. … What is changed, and changed radically, is the method of calculating probabilities. \end{quote} 

The wave function of non-relativistic quantum mechanics is most often viewed as giving rise to a probability distribution of the form $P = \Psi^{*}\Psi$. This simple interpretation, attributed to Born, actually belies a deep subtlety. Consider the case in which one has a system upon which one may perform two different measurements $a,b$ resulting in the dichotomous outcomes $a = \{a_{1},a_{2}\}$ and $b = \{b_{1},b_{2}\}$. Kolmogorov theory shows that the sum of probabilities takes the form

\begin{displaymath}
P(b=\beta) = \sum_{a_{i}}P(a=a_{i})P(b=\beta|a=a_{i})
\end{displaymath}  

However, if one attempts the same calculation in a quantum mechanical setting using the Born rule then one obtains the formula

\begin{widetext}
\begin{displaymath}
P(b=\beta) = \sum_{a_{i}}P(a=a_{i})P(b=\beta|a=a_{i}) + 2\cos \theta\sqrt{P(a=a_{1})P(b=\beta|a=a_{1})P(a=a_{2})P(b=\beta|a=a_{2})}
\end{displaymath}  
\end{widetext}
instead. From this result alone it is clear that quantum probability theory is of a non-Kolmogorovian type. Indeed as noted by Khrennikov, inequalities of Bell type were developed much earlier in probability theory dating back to the time of Boole and arise in situations in which one attempts to determine correlations when it is impossible to define those correlations using a single Kolmogorov probabilty space. Khrennikov \cite{Khrennikov} comments that even Kolmogorov in his original writings on probability was more sophisticated than later writers. He writes \textquotedblleft For him (Kolmogorov) it was totally clear that it is very naive to expect that all experimental contexts can be described by a single (perhaps huge) probability space\textquotedblright \cite[pg. 26]{Khrennikov}. It may not to possible to measure all of these observables simultaneously. The issue is then whether or not a set of observables exhibits probabilistic compatibility or incompatibility, that is, whether it is possible to construct a single probability space serving for the entire family. It is certainly not true in quantum mechanics due to the noncommutative nature of the set of self adjoint operators representing quantum measurements and the presence of interference terms arising from probabilities based on Born's rule. It is also not true of classical system in general, something that has been mostly ignored (Simpson's paradox in the social sciences is an example of this). 
 
Nonlocality is not required in order to obtain those results. They are wholly dependent upon the nature of the observables being measured and whether or not a single Kolmogorov probability space can be constructed.  These observations raise questions as to whether it is absolutely impossible to have local hidden variable models at the lowest levels. 

\subsection{Failure of Additivity}

The rule of additivity is fundamental in Kolmogorov's formulation of the laws of probability, providing one of its axioms. It holds, mostly, in non-relativistic quantum mechanics. Indeed suppose that a single quantum system can exist in one of a set of distinct energy eigenstates $\Psi_{i}$. These energy eigenstates form an orthonormal set of functions in some Hilbert space. Suppose that the system is now created in a linear superposition of these energy eigenstates as is permitted by the Schrodinger equation. The wave function for this superposition will take the form $\Psi = \sum_{i} w_{i}\Psi_{i}$ where the weights $w_{i}$ are chosen so that $\sum_{i} |w|^{2}_{i} =1$ which ensures that $\Psi$ can be interpreted in its own right as a probability distribution. This guarantees that $< \Psi^{*}|\Psi> = \sum_{ij} w^{*}_{i}w_{j}<\Psi^{*}_{i}|\Psi_{j}>=\sum_{i} |w|^{2}_{i}=1$.

When the energy of such a quantum system is measured it will yield a single value corresponding to one of these energy eigenstates. If the system is subjected to repeated measurements of its energy it will remain in the same energy eigenstate. This is considered due to the collapse of the wave function that occurs as a result of the measurement process. If multiple identically created copies of the system have their energies measured then these energies will be distributed according to the probability distribution given by $(w^{2}_{1},\ldots, w^{2}_{n},\ldots)$. 

The expectation value of the energy is given by $\hat E = \sum_{i} w^{2}_{i}E_{i}$. Suppose though that one asks a slightly different question, namely, fix some region of space, say $R$, and ask what is the expectation value of the energy over the region $R$. This is calculated as $\hat E_{R} = \int_{R} \sum_{i}\sum_{j} w^{*}_{i}w_{j} \Psi^{*}_{i}E_{j}\Psi_{j} dV = \sum_{i}\sum_{j} w^{*}_{i}w_{j}E_{j} \int_{R} \Psi^{*}_{i}\Psi_{j}dV$. Rewriting yields $\hat E_{R} = \sum_{j} E_{j} \sum_{i}w^{*}_{i}w_{j}\int_{R}\Psi^{*}_{i}\Psi_{j} dV$. One sees that the new probability associated to each energy $E_{i}$ is no longer $w^{2}_{i}$ but rather the more complicated $\sum_{i}w^{*}_{i}w_{j}\int_{R}\Psi^{*}_{i}\Psi_{j} dV$. This is due to the fact that the wave functions may overlap on $R$. It is only in the context of the entire space-time that the wave functions are orthogonal. There is no guarantee that $\sum_{i}\sum_{i}w^{*}_{i}w_{j}\int_{R}\Psi^{*}_{i}\Psi_{j} dV = 1$ so these probabilities are not additive even though the events, namely the $E_{i}$ cover the range of possible energy values.

Also note that in constructing a superposition state one is in essence constructing a sum of probabilities for if $\lambda_{i}$ is the eigenvalue associated with eigenstate $\Psi_{i}$ then the probability based upon the wave function of a superposition becomes
\begin{displaymath}
\Psi^{*}\Psi = [\sum_{i}P(\Psi_{i})P(\lambda_{i}|\Psi_{i})]+\text{interference terms} 
\end{displaymath}
This problem is frequently considered to be a feature of quantum mechanics because the quantum mechanical formalism allows for the phenomenon of quantum interference. That it appears in the classical realm as well is illustrated by the following simple model. Most everyone is familiar with the Danish children\textquoteright s toy, LEGO. Typical LEGO pieces are blocks of plastic having tiny solid cylinders protruding on the top surface of the block and corresponding cylindrical tubes in place on the undersurface. There are plates that can be used for mounting LEGO block structures. Consider the following scenario. There is a $2 \times 2$ mounting block fixed inside a sealed box. Within the box is a bag containing  a $1 \times 1$ block and a $2 \times 2$ block. There is dial on the outside of the box which reads 0,1,2. When the dial is set, a reading is taken of the plate and a light turns on corresponding to whether there is no block on the plate (0), a $1 \times 1$ block, whether alone or combined with a $2 \times 2$ block (1) or a $2 \times 2$ block again alone or in combination with a $1 \times 1$ block (2). The examiner cannot look in the box and in fact has no knowledge of the contents of the box. They can only switch the dial and note whether or not a light appears. In another room a researcher can remotely arrange whatever they like on the plate: no block, a $1 \times 1$, or a $2 \times 2$ block and they change the arrangement immediately following each observation of the examiner. Clearly the probabilities of no block, a $1 \times 1$ block or a $2 \times 2$ block are all 1/3. Therefore for the examiner the probabilities of obtaining a light for 0,1,2 are all $1/3$.

Now let us change the game slightly. The researcher is now permitted to take no action, place a $1 \times 1$ or a $2 \times 2$ block on the plate, or to couple the $1 \times 1$ block to the top of the $2 \times 2$ block and affix this to the plate. Setting the dial to 1 or 2 results in a light so long as the corresponding block is present regardless of whether it is alone or in combination. Note that it is impossible in this arrangement to measure for 1 and 2 simultaneously. Now what is the probability of there being a light on 1?  This probability is $1/2$ because there is a $1/4$ probability of there being a single $1 \times 1$ block and a $1/4$ probability of there being a $1 \times 1-2 \times 2$ combination. The same holds for the probability of a light on 2, while the probability of a light on 0 remains $1/4$. Note that now $P(0) + P(1) + P(2) = 1/4 + 1/2 + 1/2 = 1 1/2$. As far as the examiner is concerned, the outcomes are disjoint but the sum is not additive to 1.

A standard argument to correct this problem is to assert that the space of alternatives has been incorrectly constructed. If the examiner is allowed to look at the blocks then they might argue that only the global configurations constitute allowable events and these decompose into four equal probabilities, and then the probabilities of occurrence of the individual smaller blocks can be determined using conditional probabilities as per the Kolmogorov scheme. In such a case the probability of a 1x1 block becomes: $P(1\times 1)=P(0)P(1\times 1|0) + P(1\times 1)P(1\times 1|1\times 1)+P(1\times 1+2\times 2)P(1\times 1|1\times 1+2\times 2)+P(2\times 2)P(1\times 1|2\times 2)$
=1/4$\times$ 0 + 1/4$\times$ 1 + 1/4$\times$ 1 + 1/4$\times$ 0 = 1/2, 
which is the result given above. But this is a mathematical cheat because it assumes knowledge that the examiner does not and cannot possess. From the point of view of the examiner the space of alternatives was correctly constructed and they are disjoint. However they must also accept the necessity to introduce an interaction term, or to accept a non-standard form for the calculation of the total probability, namely $P(total)= P(0)+P(1)+P(2)+I(0,1,2) = 1/4 + 1/2 + 1/2 -1/4$.

Arguing that this scenario is contrived is also a cheat because this is precisely the situation for the experimental physicist. Measurement devices provide only the results of measurements, they do not yield the states of the systems being measured which cannot be directly observed. The idea of an particle being in a superposition comes out of theory, not direct observation. As in many cases experiments are contrived to create a collection of particles in a pre-determined state so that the examiner has some knowledge beforehand. If no such knowledge is obtainable, or if simultaneous measurements cannot be made it may not be possible to confirm the existence of such interaction states so as to expand the space of alternatives in such a manner so as to preseve the Kolmogorov property. The preservation of Kolmogorovian probability appears to require that one begin with the most basic \textquoteleft natural kinds\textquoteright $\:$ from which all other functions are derived, but if we do not know that combinations exist we can only deal with the event set in hand. 

Interference creates a failure of the usual additivity in the quantum mechanical case and in this classical case as well. Thus one must accept that the Kolmogorov axioms may work well in many circumstances but there may be other situations in which they fail, and instead of denying the validity of these alternative situations, we should embrace the idea that, just as in the acceptance of non-Euclidean geometry, we should accept the idea of non-Kolmogorovian probability theories.  There is nothing a priori wrong with the question that the examiner asks, nor the interpretation made of the conditions under which the question is to be answered, unless one requires that the answer follow the conditions of Kolmogorovian probability theory. Instead this very simple example urges us to accept the existence of non-Kolmogorovian probability theories, even in the classical setting, and in situations in which the basic elements of observations are derived from prior conditions that are able to interact or superpose in some manner. Such possibilities are abundant in quantum mechanics but also in the life and social sciences. Khrennikov has emphasized this point in his extensive writings on non-Kolmogorovian probability theory \cite{Khrennikov}. 
The problem arises in this example because the Lego pieces are able to interact. The problem arises in quantum mechanics because in a superposition state the individual eigenstates interact. There is no fundamental difference between these two cases. Kolmogorov theory presumes that there is no interaction between individual events and that distinct events correspond to distinct natural kinds. This example might seem trivial yet it lies at the heart of the problem of measurement. When we consider an electron, for example, in a superposition of distinct energy states, what exactly do we mean? When we ask the question of its energy, we are asking exactly the question of the examiner above – whether or not when we observe the electron do we observe one of its supposed constituent energy states. We do not think of the superposed electron as a different natural kind from the non-superposed electrons. Rather we think of an electron in a particular state, and that very same electron can change state into one of the energy eigenstates or back into a different superposed energy state. The electron is the natural kind, not the state. Moreover an electron, to the best knowledge available today, does not appear to be composed of smaller natural kinds, it is a single whole. 

The point to be made is that in any model providing a realist interpretation of quantum mechanics it is necessary to pay close attention to the subtle nature of interactions among the various elements that make up the model. One must be very careful not to project fundamental features of Kolmogorovian probability theory onto non-Kolmogorovian probability theories. These are subtle conceptual and logical errors which I suspect have arisen time and time again in our attempts to understand quantum mechanics. Over the past century we have become comfortable with non-Euclidean geometry and such logical errors no longer plague the field. Hopefully the same may one day be true of quantum mechanics. The most important consideration in constructing an alternative model of quantum mechanics is to ensure that the non-Kolmogorovian nature of the probabilities be preserved in the model.

There are other issues at play besides the type of constitutional interference as noted above. The inability to construct a single space upon which all of the probability functions can be constructed is another feature that is frequently ignored, even in the classical application of Kolmogorovian probability theory. This problem arises when one has a collection of distinct suitable state spaces upon which Kolmogorovian probabilities are developed and then one attempts to combine these into a single space in order to calculate correlations and conditional probabilities and still expect the original individual probabilities to be derivable. Probability theorists have known for a century that such a construction is not always possible and yet time and again researchers proceed as if they can carry out such a construction.

In the model to be constructed below we shall utilize combinatorial games with tokens which frequently give rise to non-Kolmogorovian probability structures.

\subsection{Failure of Stationarity}

Let us consider another classical example. Consider the following iterated function system, denoted $\phi$. Consider a simple $2 \times 2$ block into which we place different numbers. For example one might configure the block as $\begin{array}{cc}
1 & 2 \\
4 & 3 \\
\end{array}$. There are two transformations $\alpha,\beta$ which can be applied to such a block. Transformation $\alpha$ interchanges the elements in the first row. Thus $\alpha(\begin{array}{cc}
1 & 2 \\
4 & 3 \\
\end{array}) = \begin{array}{cc}
2 & 1 \\
4 & 3 \\
\end{array}$ while transformation $\beta$ interchanges the elements in the second row, so $\beta(\begin{array}{cc}
1 & 2 \\
4 & 3 \\
\end{array}) = \begin{array}{cc}
1 & 2 \\
3 & 4 \\
\end{array}$. Now consider an iterative function system defined on the space of blocks $\{\begin{array}{cc}
1 & 2 \\
4 & 3 \\
\end{array},\begin{array}{cc}
2 & 1 \\
4 & 3 \\
\end{array},\begin{array}{cc}
1 & 2 \\
3 & 4 \\
\end{array},\begin{array}{cc}
2 & 1 \\
3 & 4 \\
\end{array}\}$. For simplicity denote this set as $\{a,b,c,d\}$ respectively.

Applying $\alpha,\beta$ to this set induces the following transformations where a move to the right represents an application of $\alpha$ and a move down represents an application of $\beta$.

\begin{displaymath}
\begin{array}{ccccc}
a & \rightarrow & b & \rightarrow & a \\
\downarrow &  & \downarrow &  & \downarrow \\
c & \rightarrow & d & \rightarrow & c \\
\downarrow &  & \downarrow &  & \downarrow \\
a & \rightarrow & b & \rightarrow & a \\
\end{array}
\end{displaymath} 

If we start with $a$ and repeatedly apply $\alpha,\beta$ we will end up with a collection of possible sequences of blocks which can be represented in the form of a tree in which an arrow down and to the left means apply $\alpha$ and down and to the right means apply $\beta$.

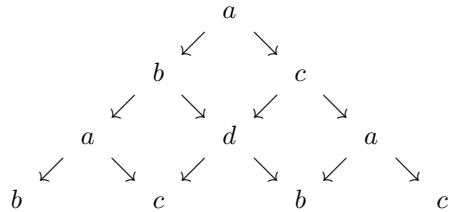
\begin{displaymath}
\begin{array}{ccccccccccccc}
 &  &  &  &  &  & a &  &  &  &  &  &  \\
 &  &  &  &  &  \swarrow&  & \searrow &  &  &  &  &  \\
 &  &  &  & b &  &  &  & c &  &  &  &  \\
 &  &  & \swarrow &  & \searrow &  & \swarrow &  & \searrow &  &  &  \\
 &  & a &  &  &  & d &  &  &  & a &  &  \\
 & \swarrow &  & \searrow &  & \swarrow &  & \searrow &  & \swarrow &  & \searrow &  \\
b &  &  &  & c & &  &  & b &  &  & & c \\
\end{array}
\end{displaymath}

Each layer represents a possible outcome after a fixed $n$ iterations. In order to determine the probability of observing a particular outcome one must sum up the number of paths leading to said outcome and then divide by the total number of possible paths. Summing over the paths leading to each outcome leads to a tree diagram

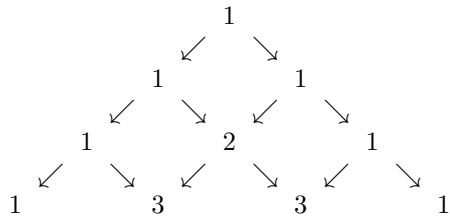
\begin{displaymath}
\begin{array}{ccccccccccccc}
 &  &  &  &  &  & 1 &  &  &  &  &  &  \\
 &  &  &  &  &  \swarrow&  & \searrow &  &  &  &  &  \\
 &  &  &  & 1 &  &  &  & 1 &  &  &  &  \\
 &  &  & \swarrow &  & \searrow &  & \swarrow &  & \searrow &  &  &  \\
 &  & 1 &  &  &  & 2 &  &  &  & 1 &  &  \\
 & \swarrow &  & \searrow &  & \swarrow &  & \searrow &  & \swarrow &  & \searrow &  \\
1 &  &  &  & 3 & &  &  & 3 &  &  & & 1 \\
\end{array}
\end{displaymath} 

 Dividing by the total number of paths at each level gives
 
 
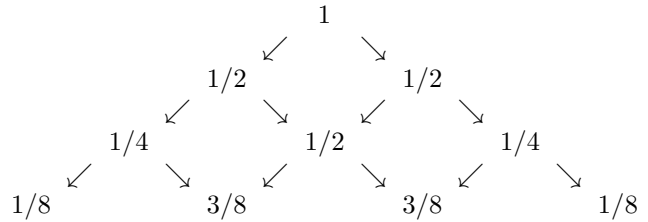
\begin{displaymath}
\begin{array}{ccccccccccccc}
 &  &  &  &  &  & 1 &  &  &  &  &  &  \\
 &  &  &  &  &  \swarrow&  & \searrow &  &  &  &  &  \\
 &  &  &  & 1/2 &  &  &  & 1/2 &  &  &  &  \\
 &  &  & \swarrow &  & \searrow &  & \swarrow &  & \searrow &  &  &  \\
 &  & 1/4 &  &  &  & 1/2 &  &  &  & 1/4 &  &  \\
 & \swarrow &  & \searrow &  & \swarrow &  & \searrow &  & \swarrow &  & \searrow &  \\
1/8 &  &  &  & 3/8 & &  &  & 3/8 &  &  & & 1/8 \\
\end{array}
\end{displaymath}

We now find the probability for a given outcome by summing over the probabilities for all paths leading to the outcome. This yields the following probability distributions:

Level $0: f=(1,0,0,0)$,
Level $1: g=(0,1/2,1/2,0)$,
Level $2: h=(1/2,0,0,1/2)$,
Level $3: g=(0,1/2,1/2,0)$.

Thus as we successively iterate the system, the probability distributions at successive times oscillate $f,g,h,g,h,g \ldots$.

It is important to note that no probability distribution has been assigned a priori to the choices of the elements $\alpha,\beta$. If a probability is pre-assigned then the above probabilities need to be modified by multiplying each path segment by the probability assigned to the particular path choice, either $\alpha$ or $\beta$.

The above model provides a simple, non-deterministic dynamical system which is entirely classical, where the probabilities are determined by a discrete version of real valued path integrals, and which yield temporally oscillating, spatially non-stationary probability distributions. The point of this example is to highlight the fact that quantum mechanical systems are not alone in having a probability structure that can be calculated utilizing path integrals. Combinatorial based classical systems such as the example described above and many combinatorial games possess this path integral structure. Whether or not there exists a limiting stationary probability distribution over the state space depends upon the tree structure induced by the dynamics of the combinatorial operations. Most iterated function systems involving actions on a continuous real space require some form of contraction so as to ensure that an invariant or stationary measure exists on the state space. 

\subsection{Failure of the Law of Total Probability}

Let us stay with the block space. Consider a second iterated function system acting on the same space of blocks. Call it $\rho$. This time we have a single function $\gamma$ acting on the block space. The action of $\gamma$ on any block is to interchange the first and second columns. That is $\beta(\begin{array}{cc}
1 & 2 \\
4 & 3 \\
\end{array}) = \begin{array}{cc}
2 & 1 \\
3 & 4 \\
\end{array}$.

The action of $\gamma$ on the block space is simple: $a \rightarrow d$ and $b\rightarrow c$. 

Again starting with block $a$ and repeating the procedure of the previous section yields the following probability distributions:

Level $0: f=(1,0,0,0)$,
Level $1: j=(0,0,0,1)$,
Level $2: i=(1,0,0,0)$,
Level $3: j=(0,0,0,1)$.

Note that the distribution for $\rho$ is distinct from $\phi$ in the previous example and that these represent two distinct iterated function systems. Now let us consider the iterated function system $\sigma$ on the block space generated by $\{\alpha,\beta,\gamma\}$. 
Again start with block a. Applying either of $\alpha, \beta$ or $\gamma$ yields the outcomes $b,d,c$. Applying the maps to these outcomes yields outcomes
$a,c,d,c,a,b,d,b,a$. Outcomes are repeated in the above listing as each represents a distinct path down the tree. Applying the maps once more yields the 27 outcomes $b,d,c,d,b,a,c,a,b,d,b,a,b,d,c,a,c,d,c,a,b,a,c,d,b,d,\!c$. The probability distributions are thus 

Level $0: f=(1,0,0,0)$,
Level $1: k=(0,1/3,1/3,1/3)$,
Level $2: l=(1/3,2/9,2/9,2/9)$,
Level $3: m=(2/9,7/27,7/27,7/27)$.

We may denote this iterated function system as $\sigma=(2/3)\phi + (1/3)\rho$. If we combine their probability distributions then we would obtain  

Level $0: f=(1,0,0,0)$,
Level $1: 2/3g + 1/3j=2/3(0,1/2,1/2,0) = 1/3(0,1,0,0) = (0,2/3,1/3,0)$,
Level $2: 2/3h + 1/3i =2/3(1/2,0,0,1/2)+1/3(1,0,0,0)=(2/3,0,0,1/3)$,
Level $3: 2/3g + 1/3j=2/3(0,1/2,1/2,0)+1/3(0,1,0,0)=(0,2/3,1/3,0)$.

Note that $k\neq 2/3g + 1/3j$, $l\neq 2/3h+ 1/3i$ and $m\neq 2/3g + 1/3j$. Thus although these probabilities should add according to the usual notions of probability theory they do not because there is an interaction effect. In this case, although $\phi$ and $\rho$ are distinct iterated functions systems and their superposition gives rise to a perfectly good iterated function system, the resulting probability distribution functions cannot be obtained from a simple weighted sum of the individual prior probability distributions because there is an interaction, namely $\alpha\beta = \gamma$.

This is a simple example but it bears a formal similarity to the situation in quantum mechanics where one considers linear superpositions of eigenfunctions. In both cases, difficulties arise with the usual composition of probability distribution functions because of interaction effects, usually function overlap in the case of quantum mechanical systems, algebraic effects in the simple iterated function system discussed here. The significance of this example is that this demonstrates the failure of additivity even in the case of a classical system with real valued functions. Quantum mechanics is not necessary for such non-Kolmogorovian effects to appear. 

\subsection{Failure of Bell's Theorem}

Let us now consider the following pair of single player combinatorial games. They are not very interesting as games but they illustrate a feature of games which is that they can defeat the Bell inequality under certain conditions. For this example consider a pair of games played out on the previously defined $2 \times 2$ blocks $a,b,c,d$, one game using the transformation $\alpha$ and the other $\gamma$. We consider sequential game play, and we are interested in the outcome following every two steps of play. As described previously, Bell's original theorem involves relationships among three correlation functions based upon spin measurements on a pair of entangled particles. In this example we consider correlation functions based on trajectories defined by repeated game play, with different initial conditions replacing the different orientations of measurement.

We consider three initial conditions, $a,b,d$. In the correlations defined below, the first variable refers to the game generated by $\alpha$ and the second to that generated by $\gamma$. Measurements of $a,b,c,d$ have defined values of $1,1/2,-1/2,-1$ respectively.

Play using $\alpha$ or $\gamma$ yield distinct trajectories. Nevertheless when we restrict ourselves to two play games we note that $\alpha\alpha =id$ and $\gamma\gamma = id$ so that we always obtain constant trajectories, namely just the initial condition.

Bell's inequality takes the form

\begin{displaymath}
1 + E(\mathbf{b},\mathbf{c}) \geqq |E(\mathbf{a},\mathbf{b})-E(\mathbf{a},\mathbf{c})|
\end{displaymath}

\noindent Note that only the expectations values are important, not the circumstances under which they were generated. It is only important that the measurement values in directions $\mathbf{a},\mathbf{b}$ be $\pm 1$. So long as we ensure that under two conditions the measurement values also be $\pm 1$ then we meet the essential mathematical requirements of Bell's Theorem. Clearly in this simple example if we choose initial conditions $a,d$ then we shall obtain measured expectation values of $1,-1$ respectively. Choose for the third initial condition the block $b$. Bell's inequality takes the form

\begin{displaymath}
1 + E(d,b) \geqq |E(a,d)-E(a,b)|
\end{displaymath}

Since the play corresponds to simply applying the identity to the initial condition, the probability of observing each initial condition is 1 as is the probability of observing the pair of initial conditions, so that the expectation value of the measurement of the product becomes simply the product of the measurements. Therefore calculations of these correlations yields

\begin{displaymath}
1 + (-1)(1/2) \geqq |(1)(-1)-(1)(1/2)|
\end{displaymath}

or

\begin{displaymath}
1 -1/2 = 1/2  \geqq |-1-1/2| =|-3/2|=3/2
\end{displaymath}

\noindent which is clearly false. Thus we have a simple discrete, classical system which nevertheless exhibits correlations that violate the Bell inequality.
The violation of the inequality holds for this special triple of initial conditions just as the violation of Bell's theorem in quantum mechanics occurs for certain measurement directions.

This model is intentionally simplistic. The point is to demonstrate that the assumption that a classical dynamical system \emph{must} be describable by a Kolmogorov type probability theory is not actually correct. This example, while involving simple one player games, may also be understood as a deterministic dynamical system. As such it is deterministic and local and there is no interaction between the two systems. The coupling arises because of the choice of initial conditions and the processes themselves. The coupling is not at the level of the individual events but rather at the level of the dynamics generating those events. It also demonstrates that the conclusion from Bell's theorem that only a deterministic nonlocal hidden variable theory is capable of describing quantum mechanical phenomena is not necessarily true. This observation is in keeping with Palmer, who showed that an iterated function system may reproduce quantum mechanical spin statistics while still avoiding Bell's theorem. Palmer gets around Bell by showing that the necessary correlation functions fail to exist. This simple example shows that the theorem may be defeated directly. In situations in which the dynamics is generated by games (and possibly iterated function systems as well), the probability structure need not be Kolmogorov and consequently it may be possible to defeat the Bell inequality. 

Khrennikov summed up these insights, stating \textquotedblleft Violation of Bell's inequality is merely an exhibition of non-Kolmogorovness of quantum probability, i.e. the impossibility of representing all quantum correlations as correlations with respect to a single Kolmogorov probability space\textquotedblright \cite[pg. 6]{Khrennikov}. Khrennikov has developed these ideas of non-Kolgorovian probability in his V\" axj\" o model of contextual probability theory. The details of this approach are not necessary here but it generalizes the addition formula for quantum probabilities and applies this to classical events. What is important is that the most fundamental assumption of Bell that classical events must follow the rules of Kolmogorovian probability theory is not true in general and so the conclusions derived from Bell's theorem related to the necessity of nonlocality in any hidden variable model of quantum mechanics are also not universally valid. Inspired by these ideas we turn now to a set of mathematical approaches which capture this idea of non-Kolmogorovness in classical settings and so open the door to realist quasi-local hidden variable models of quantum mechanics.

\section{Process Theory}

Process is a construct well recognized in psychology and biology, for example the emergence theories of Trofimova \cite{Trofimova} and Varela \cite{Varela}. In physics the notion of process generally refers to an interaction between entities that unfolds in time. Processes may change certain dynamical parameters associated with continuous symmetries such as energy, position or momentum. Examples of these include scattering, state transitions, capture and emission. Processes may change certain discrete parameters associated with intrinsic characteristics like charge, charm, strangeness, lepton or baryon number. Creation, annihilation and decay are examples of these. Conceptually, fundamental physical entities enter into or emerge from various processes but their very existence is not viewed as arising out of process.

The view of Whitehead stands in marked contrast. Whitehead views reality as emerging out of a lower level of reality consisting of \emph{actual occasions}. Fundamental physical entities are viewed as emergent configurations of actual occasions. An analog lies in attempts in the 1980's to model reality as a cellular automaton where particles appeared as patterns manifesting over time on the cellular automaton lattice \cite{wolfram}. Process theories, particularly the theory of Whitehead \cite{Whitehead}, possess several essential features that need to be considered in creating a representational system that expresses them.

\begin{enumerate}
\item  The basic elements of experience, actual occasions, have a richer character than is generally attributed to elements of reality. Actual occasions possess a dual character. On the one hand they form a fundamental component of the fabric of reality. On the other hand, they serve as information for the creation of subsequent actual occasions. 
\item  Process theory is a generative theory. The actual occasions that form the essence of reality come into existence through a process of prehension, in which the information residue of prior occasions is interpreted and new occasions generated creatively in a non-deterministic manner. 
\item  Actual occasions are transient in nature. They arise, linger briefly and then fade away. In contrast to current physical thinking, process theory asserts the existence of a transient \textquoteleft now\textquoteright.
\item  Essential to process is the idea of becoming. This is subtly different from the notion of generation. For example, an iterated function system generates a trajectory by the repeated application of the function to a previous point: $x$, $f(x)$, $f(f(x))$,… However, the space upon which this function acts exists a priori. A trajectory in the space is generated, the space itself is not. In process theory, the space itself does not exist a priori, indeed it does not exist at all except as an idealization in some mathematical universe. All that exists is a collection of actual occasions that are continually in the process of becoming. An actual occasion has no existence unless and until it is brought into existence through the action of prehension. It subsequently fades into non-existence and any future influence that it might have arises solely through its representation in some form of memory.
\item  In process theory events are fundamentally discrete, being comprised of vast numbers of actual occasions. The perception of events as being continuous may well be a deeply ingrained illusion. Its abstraction in mathematical form has given rise to powerful analytical tools, so much so that continuity has been reified as a property of space-time and histories and entities. Process theory asserts that entities and their motion are actually discrete and their apparent continuity is again a consequence of the process of idealization inherent in the formation of an interpretation.
\item Actual occasions are held to be holistic entities. It may be convenient for conceptual, descriptive or analytical purposes to consider actual occasions as consisting of individual \textquoteleft parts\textquoteright, but this again constitutes an idealization or contrivance. Each actual occasion must be considered to be a whole unto itself and any information or influence attributed to an actual occasion must be attributed to the actual occasion as a whole and not to any of its supposed parts. Any such parts  must be considered to be unobservable as must any presumed properties or characteristics of these parts. Properties and characteristics that may be observed by other actual occasions must be attributed solely to the actual occasion as a whole.
\end{enumerate}

From a process perspective all matter would be viewed as emergent, arising from the evolution of actual occasions. These actual occasions would not be accessible to material entities in much the way that mind is incapable of sensing the actions of individual neurons, even though mind is emergent from the actions of neurons. Being emergent, the laws governing the behaviour of actual occasions need not be those of quantum mechanics, though the behaviour of entities emerging at the lowest spatiotemporal scales should obey those laws. Likewise, entities emerging from these fundamental quantum entities need not obey the laws of quantum mechanics, or at least quantum mechanical laws need not be relevant for understanding their behaviour, just as the laws governing the action of mind are not the same as the laws governing the neurons giving rise to it. This is a common situation in the theory of emergence. Although lower level entities may give rise to higher level entities, the relationship between these two may be such that there is no one-one correspondence between the behaviours at one level and those at the other level, so that the laws governing the lower level become irrelevant for understanding behaviour at the higher level.

There are two other aspects of process that deserve mention. First of all, process has an inherently non local character. A biological organism is an expression of a vast array of processes but these processes are local only in a naive and superficial sense. The entities that participate in these processes are distributed widely in space and time. Mental processes cannot even be localized to the brain as the body and the environment play important roles. Secondly, the entities that participate in process are often fungible. Although a whole organism may not be fungible, its constituent molecules most certainly are, and sometimes components are not even of the same species as the organism, particularly in the case of digestive processes.  A board game such as Chess is a simple example of a process. Although individual moves of chess pieces are local, the choice of which piece to move on a given play is inherently non local, though certain game positions may favour local or non local choices. Chess is fungible, so long as any replacement respects the current arrangement of pieces. Chess can be played anywhere, at any time, with almost any objects, real or virtual, so long as a suitable correspondence is established between the objects and their movements and their roles as chess pieces. Processes in themselves exist in an abstract, aspatial and atemporal world, while the actual occasions that they generate manifest in space and time and bear specific relations to one another that are interpreted as properties.

\section{Process Interpretation of the Wave Function}

There are subtleties of dynamics that are not easily captured by the standard functional analytic formulation of quantum mechanics. Consider the issue of being \textquoteleft bound \textquoteright. Classically it is a fairly straightforward matter to determine whether a particle is bound to another particle because the trajectory of the bound particle will form a closed path with the binding particle lying in the interior of the path. In quantum mechanics this is not so straightforward since particles do not follow trajectories that can be mapped. A free or a bound particle can, in principle, be found anywhere in space.  Its mere detection says little about its dynamical state. A detailed determination of the shape of the wave function would help but is unfeasible. Moreover a free particle could be stationary and have a spherically symmetric wave function just like a bound particle. The main difference is that the probability of the free particle being near the centre is fairly large while for the bound particle is it fairly small. A free particle can propagate but how does one distinguish the random motion of a bound particle and the propagation of a free particle, when both can appear more or less anywhere at any time? Moreover, in the case of a spherical potential, the potential extends throughout all of space and so in considering when a free particle becomes bound it is not clear when exactly one is to apply the bound equation and not the free particle equation. In principle the free particle could become bound anywhere and at any time. And if it is bound to one particle could it not also become bound to another particle? To every other particle? The equation describing the dynamics of the particle must change between free and bound conditions and so an additional consideration must come into play to determine when this takes place. What is this additional consideration and how does a particle \textquoteleft know\textquoteright$\;$ when to apply it?

The classical interpretation of the wave equation is that it provides a probability distribution for the position of the quantum system. More precisely it provides a probability distribution for a detection by a position measurement apparatus, said detection usually attributed to the presence of a particle in that location at that time. This is not a problem given an ensemble view of the wave function, whereby repeated measurements of an ensemble of identically created particles are conducted and the probability distribution of those measurements calculated. There are problems, however, when one wishes to attribute the probability distribution to a single particle or to attribute a physical reality to the wave function much as one attributes reality to the electromagnetic wave function. In the case of the latter there are demonstrable effects having an electric and magnetic character which can be attributed to the electromagnetic wave so its reality is not really questioned anymore. The Schr\"odinger wave functions are quite different in character. Although Aharonov and Vaidman \cite{Aharonov} have suggested that the wave function of a single particle could be detected using quantum non-demolition measurements, it is only certain statistical measures that can be detected, not the wave function itself.

Consider again the situation of a particle in a spherical potential, this time in the bound state. If the particle is in an eigenstate of the Hamiltonian, say being in energy level $n$, with angular momentum $l$ and spin angular momentum $m$, then the wave function takes the form

\begin{displaymath}
\Psi_{nlm}(r,\theta,\phi) = Ae^{-\frac{i}{\hslash}E_{n}t}R_{nl}(r)P^{m}_{l}(\cos\theta)e^{im\phi}
\end{displaymath}

\noindent where $R_{nl}(r)$ is the radial wave function (real valued), $P^{m}_{l}$ is the associated Legendre polynomial (real valued) and $A$ the normalization constant.

The probability distribution for this particle is given by

\begin{multline*}
\Psi^{*}_{nlm}\Psi_{nlm}=A^{2}e^{\frac{i}{\hslash}E_{n}t}R_{nl}(r)P^{m}_{l}(\cos\theta)
 e^{-im\phi}\times\\
 e^{-\frac{i}{\hslash}E_{n}t}R_{nl}(r)P^{m}_{l}(\cos\theta)e^{im\phi}=\\ A^{2}R^{2}_{nl}(r)(P^{m}_{l}(\cos\theta))^{2}
\end{multline*}
 
Now the Hamiltonian in this case is time independent and so one would expect that the probability distribution would also be time independent and that is indeed the case. 

Consider now the case in which the particle is in a superposition of adjacent energy levels. The wave function is this case is given by

\begin{multline*}
\frac{1}{\sqrt{2}}\Psi_{nlm}+ \frac{1}{\sqrt{2}}\Psi_{(n+1)l'm'} =\\
Ae^{-\frac{i}{\hslash}E_{n}t}R_{nl}(r)P^{m}_{l}(\cos\theta)e^{im\phi}+\\
Be^{-\frac{i}{\hslash}E_{n+1}t}R_{(n+1)l'}(r)P^{m'}_{l'}(\cos\theta)e^{im'\phi}
\end{multline*}

The probability distribution in this case is given as

\begin{multline*}
P(r,\theta,\phi)=(1/2)A^{2} R^{2}_{nl}(r)(P^{m}_{l}(\cos\theta))^{2}+\\
(1/2)B^{2} R^{2}_{(n+1)l'}(r)(P^{m'}_{l'}(\cos\theta))^{2}+\\
Re\{ABe^{-\frac{i}{\hslash}(E_{n}-E_{n+1})t}R_{nl}(r)P^{m}_{l}(\cos\theta)\times\\
R_{(n+1)l'}(r)P^{m'}_{l'}(\cos\theta)e^{i(m-m')\phi}\}=\\
(1/2)A^{2} R^{2}_{nl}(r)(P^{m}_{l}(\cos\theta))^{2}+\\
(1/2)B^{2} R^{2}_{(n+1)l'}(r)(P^{m'}_{l'}(\cos\theta))^{2}+\\
Re\{ABe^{-\frac{i}{\hslash}(E_{n}-E_{n+1})t-\hslash(m-m')}R_{nl}(r)P^{m}_{l}(\cos\theta)\times\\ R_{(n+1)l'}(r)P^{m'}_{l'}(\cos\theta)\}=\\
(1/2)A^{2} R^{2}_{nl}(r)(P^{m}_{l}(\cos\theta))^{2}+\\
(1/2)B^{2} R^{2}_{(n+1)l'}(r)(P^{m'}_{l'}(\cos\theta))^{2}+\\
AB\cos(-\frac{i}{\hslash}(E_{n}-E_{n+1})t-\hslash(m-m'))R_{nl}(r)P^{m}_{l}(\cos\theta)\times\\
R_{(n+1)l'}(r)P^{m'}_{l'}(\cos\theta)
\end{multline*}

In this case even though the Hamiltonian remains time independent the probability distribution function now acquires a temporal fluctuation by virtue of an interaction term between the two eigenstates. Thus one no longer has a stationary probability distribution. However, the time average of this probability distribution is

\begin{multline*}
P'(r,\theta,\phi)=(1/2)A^{2} R^{2}_{nl}(r)(P^{m}_{l}(\cos\theta))^{2}+\\
(1/2)B^{2} R^{2}_{(n+1)l'}(r)(P^{m'}_{l'}(\cos\theta))^{2}
\end{multline*}

\noindent which is the usual probability distribution expected from combining the individual distributions. The loss of stationarity would appear to make any attempt to determine this probability distribution experimentally either difficult or impossible. Even in the case of a non demolition experiment it would be impossible to determine the distribution without synchronizing position sampling to the frequency of the fluctuation and without knowing the phase delay, both of which would require measuring the differences in energy levels and spin angular momenta between the two states which would appear to require a demolition experiment. 

If one attempted to measure the probability distribution with a single particle this would have to be done at a series of distinct times, say $t_{1}, \dots, t_{n}$. The functions being sampled at each time would differ, being 

\begin{multline*}
(1/2)A^{2} R^{2}_{nl}(r)(P^{m}_{l}(\cos\theta))^{2}+\\
(1/2)B^{2} R^{2}_{(n+1)l'}(r)(P^{m'}_{l'}(\cos\theta))^{2}+\\
AB\cos(-\frac{i}{\hslash}(E_{n}-E_{n+1})t_{1}-\hslash(m-m'))R_{nl}(r)P^{m}_{l}(\cos\theta)\times\\
R_{(n+1)l'}(r)P^{m'}_{l'}(\cos\theta),\ldots, \\
(1/2)A^{2} R^{2}_{nl}(r)(P^{m}_{l}(\cos\theta))^{2}+\\
(1/2)B^{2} R^{2}_{(n+1)l'}(r)(P^{m'}_{l'}(\cos\theta))^{2}+\\
AB\cos(-\frac{i}{\hslash}(E_{n}-E_{n+1})t_{n}-\hslash(m-m'))R_{nl}(r)P^{m}_{l}(\cos\theta)\times\\
R_{(n+1)l'}(r)P^{m'}_{l'}(\cos\theta)
\end{multline*}

If one happened to be sampling at the same frequency as the fluctuation, then one would obtain the mean distribution shifted by a systematic drift term $\cos(i\hslash(m-m'))R_{nl}(r)P^{m}_{l}(\cos\theta)
R_{(n+1)l'}(r)P^{m'}_{l'}(\cos\theta)$.

If one knew the phase delay one might offset it, obtaining the average
distribution. If one samples the times uniformly and randomly, then these fluctuations would, on average, cancel each other out, again leaving the average distribution. However, the average distribution is \textit{not} the wave function, since the fluctuating term is not simply a random variation but rather an integral part of the wave function. Indeed the mean wave function is what would be expected from Kolmogorovian probability theory, which we already know to be inconsistent with quantum mechanics.

In this case we see that the only way in which the actual wave function can be detected is if it were possible to carry out a series of quantum non-demolition experiments on an ensemble of particles, not a single particle. 
One could not simply measure the frequency with which particles appear since such a distribution would actually have to be measured over time, and thus one would not obtain the actual distribution but only a time averaged version. The actual distribution would require an ensemble of particles whose positions could be sampled simultaneously at repeated times, the frequencies being determined for each individual time. The fluctuating distribution thus has meaning only in relation to an ensemble of particles since it is only with an ensemble that it can be measured at all. It is not at all clear how attributing a probability distribution to a single particle in this case would make any sense.

These considerations suggest problems in the interpretation of the wave function, at least in so far as single particles are concerned. For the most part, treating the wave function as an expression of ensemble behaviour is consistent with experiment as well being theoretically consistent. It admits the possibility, at least in principle, of experimental verification. In the case of single particles, however, it appears no longer possible, in general, to verify it experimentally. Suppose for the moment that we consider the possibility that the probability interpretation of the wave function is a consequence of the statistical character of ensemble behaviour and that it simply does not apply to single particles. What then might the wave function represent?
 
These two considerations suggest that NRQM may indeed be incomplete and that additional features are needed. Suppose though that these additional factors arise because the two distinct aspects of ultimate reality - actual occasions and the processes that generate them - were conflated when the original mathematical framework of NRQM was developed. Formally, NRQM takes many of the features of classical mechanics, particularly its Hamiltonian formulation, and attempts to effect a translation to a slightly more general mathematics - from point set analysis to functional analysis. Process per se is not explicitly considered in the functional analytic framework. Perhaps it would be better to look for mathematical systems that are better equipped to represent process and then see whether NRQM could be derived within this setting. Indeed, Palmer has already shown that iterated function systems may reproduce many of the essential features of quantum mechanics, at least spin statistics \cite{Palmer}. In this paper the focus is upon models based on certain types of combinatorial games. 

Suppose that the wave function actually describes information about the process responsible for the generation of a single particle. Suppose further that the probability interpretation arises in an emergent manner in the context of a statistical ensemble of particles. Note that in most quantum mechanical formulas, particularly in path integral formulations and in quantum field theory, the wave function enters into the Lagrangian, usually coupled either to itself or to the wave function of another particle. Suppose, therefore, that the wave function describes some kind of \textquoteleft strength\textquoteright$\;$ of the generating process. Different processes would then couple through these different process strengths. Positional probability arises merely when a fundamental particle couples to a position measurement device, and that turn out to be a fairly basic coupling dependent upon a term of the form $\Psi^{*}\Psi$. In this sense the probability aspect is not an intrinsic feature of the wave function but rather an emergent feature arising out of the interaction between  the particle and the measurement device. Given such an interpretation, a single particle could indeed possess a physical wave function, which describes not the particle per se but rather the process that generates the events that we ultimately interpret as a particle. Contradictions arise because we attribute the wave function incorrectly to the particle rather than to the process.

Suppose for the moment that we allow the possibility that the phenomenon that we term particle is not a thing in itself but rather is an emergent manifestation of something more primitive; that a particle is generated and that the links between occurrences of a particle possess an informational aspect. Suppose further that the actual occurrences that are the direct manifestations of these processes occur on a spatio-temporal scale much smaller than that of the particles being generated, so small that they would be inherently unobservable to any usual material entity. Any such occurrence, being so small would not be resolvable, and so would appear to any material entity as a rather ill-defined or fuzzy object.  Let us further suppose that we represent this fuzzy primitive entity as a spatio-temporal transient. As a simple example, suppose we let each such transient have the functional form (in one dimension)

\begin{displaymath}
\frac{\sin(\sigma x-kn)}{(\sigma x-kn)}
\end{displaymath}

Each such occasion manifests its process. The strength of this process is given by the value of the wave function for the process attributed to the peak of the transient. Therefore at each point the process contributes a transient of the form

\begin{displaymath}
\Psi(kn/\sigma)\frac{\sin(\sigma x-kn)}{(\sigma x-kn)}
\end{displaymath}

An observable event becomes a summation over a multitude of these primitive events. Give an observation at point z, we associate a set $I_{z}$ of points of the form $kn\sigma$ such that $z$ is an element of the real interval $\tilde  I_{z}$ formed by filling in the gaps in $I_{z}$. To the point $z$ we can associate a function

\begin{displaymath}
\Psi_{z}(x)=\sum_{kj/\sigma\in I_{z}}\Psi(kj/\sigma)\frac{\sin(\sigma x-kj)}{(\sigma x-kj)}
\end{displaymath}
 
Assume a collection of observations $\{z_{1},\ldots, z_{n}\}$ such that the corresponding intervals $\tilde I_{z_{1}}, \ldots, \tilde I_{z_{n}}$ are disjoint. Then by the Shannon-Weiner-Kotel'nikov theorem, as the number of observations increases, the resulting sum of contributions will converge to a function $\Psi(x)=\sum_{z}\Psi_{z}(x)$ defined on the entire real line. This interpolated function becomes the wave function of the particle. The particle thus moves discretely but due to the small scale we only observe and interact with the interpolation - the wave. In this way a particle has both wave-like and particle-like aspects but there is no contradiction and no paradox - it is merely a question of scale.

\section{Game Theory}

\subsection{Combinatorial Games}

A combinatorial game is a mathematical abstraction of games that are commonly played in real life such as Tic-Tac-Toe, Dots and Boxes, Checkers, Chess, Go and so on. Combinatorial games involve players who carry out moves in an alternating manner in the absence of random elements and in the presence of perfect information. The end of play is generally heralded by an inability of the players to make a move, the last player able to move being declared the winner. Combinatorial games are to be distinguished from the games usually studied in economics and biology in which players may move simultaneously in the presence of complete or incomplete information, in which there may be random elements, and in which the end of play is measured relative to some optimality criterion applied across all possible game plays.

The formal theory of combinatorial games began with Sprague-Grundy in the 1930\textquoteright s but became a mature branch of mathematics in the 1970\textquoteright s with the work of John H. Conway and others \cite{Conway}. A close cousin, the Ehrenfeucht-Fraisse game, has been used extensively in mathematical logic and model theory to construct representations of formal systems. The focus here is on Conway\textquoteright s theory, which has its most developed expression in the study of short determinate two player partisan games, though research continues to expand the theory to long indeterminate multi-player games with generalized outcomes. Short deterministic two player partisan games form a partially ordered Abelian group. Moreover, there exists a subgroup of such games that can be interpreted as numbers and constitute the expanded field of surreal numbers. The same group admits additional elements that have an interpretation as infinitesimals, extending the field to include elements of non-standard analysis.

In the combinatorial games discussed below, it is assumed that there are two players, Left and Right, who move alternately, possess possible moves that are distinct from one another, and possess complete information about the state of the game during any play. Moreover, the nature of the game is such that play is guaranteed to end after a finite number of plays. The state of the game at any play can be fully determined. The options for a given player from a particular state of the game are simply the set of all states of the game that can follow a single play of the game by that player. There will be a distinct set of Left options and of Right options. Go is an example of such a game. The definition of a combinatorial game does, however, include one additional assumption, which is that the last player to play wins. Many games do not satisfy these conditions but are still capable of analysis within this framework. The particular definition used was chosen because of its generality and the depth of its mathematical results, but there have since been many generalizations to include transfinite play, loopy play, mis\`ere play, play with different outcome determinants, and multiple players.

The play of a game begins with some initial state (position or configuration). A player moves, resulting in a new position. The other player then moves, again resulting in a new position, and the process repeats. The complete play of the game is thus described as a (finite) sequence of such positions, terminating when no further play is possible. For convenience we can catalogue all possible sequences of game play by constructing a \emph{game tree}. Denote the players as Left and Right. Starting from a particular position, we arrange below and to the left, all possible positions that can be achieved by a move on the part of Left. Similarly we arrange below and to the right, all possible positions that can be achieved by a move on the part of Right. The process is then repeated for this new level of game and so on until no more positions can be achieved. A particular complete play of the game will correspond to a path down the game tree, beginning with the initial position and then proceeding to successive positions by alternating along left and right steps. 

The formal definition of a combinatorial game is conceptually quite confusing at first but it possesses great generality. It is inherently recursive and most constructions in combinatorial game theory arise through some (implicit) form of bootstrapping or through top-down induction. The technique is powerful and worth the mental effort to master it. Since each game begins with a game position we can define a game by that initial position. Play then becomes a sequence of games rather than a sequence of positions. Moreover each position has associated with it a specific set of positions obtainable by a move of Left, termed Left options, and another specific set of positions obtainable by a move of Right, termed Right options. Since the subsequent play of the game will depend only upon these sets of options a position may just as well be equated with these two sets of options. Therefore we associate any position $P$ with its set of Left options $P_{L}$ and its set of Right options $P_{R}$. Now for the confusing part. Taking each position to be a game, each set of options can be viewed as a set of games. Hence we define a game $G = \{G_{L} | G_{R}\}$ where $G_{L}$ and $G_{R}$  are sets of games. A game has an alternative definition in terms of its game tree, which represents the possible moves for each player from any given game position. 

The fundamental theorem of combinatorial games states that given a game $G$ with players $L$ and $R$ such that $L$ moves first, either $L$ can force a win moving first, or $R$ can force a win moving second, but not both.

This results in four distinct outcome classes for games.

These are
\begin{enumerate}
\item      Positive: $L$ can force a win regardless of who goes first
\item      Negative: $R$ can force a win regardless of who goes first
\item      Zero: The second player to play can force a win
\item      Fuzzy: The first player to play can force a win
\end{enumerate}

Positive, negative and zero games form the class of surreal numbers under addition as defined below. The fuzzy games form the class of infinitesimals. Positivity-negativity is a symmetry operation given by reversal of the roles of Left and Right. For non-partizan games (the Left and Right options are always the same), there are only two outcome classes, fuzzy and zero.

The formal definitions are \cite{Conway}:

\begin{enumerate}
\item      A combinatorial game $G$ is given as $\{G_{L} | G_{R} \}$ where $G_{L}$ and $G_{R}$ are sets of games
\item      The sum of games $G + H$ is defined as $\{G_{L} + H, G + H_{L} | G_{R} + H, G + H_{R} \}$
\item      The negative of a game G, is defined as $-G = \{-G_{R} | -G_{L} \}$
\item       For two games $G, H$, equality is defined by $G=H$ if for all games $X$, $G + X$ has the same outcome as $H + X$
\item      For two games, $G, H$, isomorphism is defined as $G \approx H$ if $G$ and $H$ have the same game tree
\item      For two games, $G, H$, we say that $G \geq H$ if for all games $X$, Left wins $G + X$ whenever Left wins $H + X$
\item      A game $G$ is a number if all elements of $G_{L}$ and $G_{R}$ are numbers and $g_{L} < G < g_{R}$  for all $g_{L} \in G_{L}$ and $g_{R} \in G_{R}$ .
\end{enumerate}

For any integer $n$, a game in which Left has $n$ free moves is assigned the number $n$, while any game in which Right has $n$ free moves is assigned the number $-n$. In the case of short games, the number assigned will be a dyadic rational, i.e. an integer of the form $m/2n$  for some integers $n,m$. The number of the sum of two games that are numbers is the sum of the numbers of the individual games. Multiplication and division can be defined on games that are numbers in such a way that these games form a field, the surreal numbers. Surreal numbers that are non dyadic rationals arise through a consideration of games of transfinite length and include the rationals, the reals and the ordinals. They may be generated using techniques similar to that of Dedekind cuts for the creation of the reals.

Games can be generated recursively starting with the simplest game $0 = \{ | \}$. This is the game with no options at all. Call this day 0. At day 1, one may construct four possible games, $\{ | \}, \{ | \{ | \}\}, \{\{ | \} | \}$, and $\{\{ | \} | \{ | \}\}$ denoted 0, -1, 1 and * respectively. There are 36 games at day 2, 1474 games at day 3 and at day 4 somewhere between $3 \times 10^{12}$  and $10^{434}$ games \cite{Albert}. 

If tokens are added to these games then it becomes possible to form a vector space, and with a suitable notion of commutation, a Lie algebra. The significance of this it that is now becomes possible, at least in principle, to use token combinatorial games as representations of Lie algebras, and thus of the fundamental processes of nature vis-\`a-vis their description in the standard model of quantum field theory.

\subsection{Game sums and products}

One distinct advantage of combinatorial game theory is that it admits many different kinds of linearity, that is, many different kinds of sum may be defined. In every case, however, a key feature of a sum is that on any given play the player whose turn it is may play either in $G$ or in $H$. Thus play alternates between the two games but not necessarily in a sequential manner. The combinatorial or disjoint sum defined in the previous section describes games that may be thought of as being played out on different boards. The interpretation of $G + H$ is that on a given play a move in one game will have no effect upon play in the other game. Let us assume, however, that the distinct games are being played out on the \emph{same} board, albeit with different pieces or tokens. The disjoint sum may still be defined in this case so long as play of one game has no effect upon play of the other game. An example might be where tokens from different games may be applied to a single location on the board without affecting one another in any way. Another occurs when play is localized to non-overlapping regions of the board. In physics an example of such a situation is in the dynamics of bosons, where multiple bosons may occuy the same spatio-temporal location without affecting one another in any manner. 

We may define an exclusive sum of $G,H$ denoted $G \hat + H$. In the exclusive sum, no move of $G$ may occur on a site occupied by a piece of $H$ and vice-versa, otherwise there are no restrictions on game play. There is a weak kind of interaction present between these two games but play of one game is not determined by the other, merely constrained sometimes and at some locations. It is a rather passive kind of interaction and there is no real interchange of information between the two games. Physically speaking, there is no exchange of energy between the games. An example in physics of such a situation is in the dynamics of fermions, which are not allowed to occupy identical states. 

A third sum may be define when certain moves of $G$ influence which moves of $H$ may subsequently be played and vice-versa. In this case, information from the play of one game has an effect upon the subsequent play of the other game and so an exchange of relevant information indeed takes place. Play may or may not be exclusive. In physics, the situation of a binding of two particles would be describable by such a sum. We denote an interactive sum by $G \oplus H$, but understand that this does not define a fixed form of game play. Rather, its interpretation will depend upon the particular games and their context. At best one can say that it will be an element from a set of possible forms of game play. Note that technically an exclusive sum is really a form of interactive sum but it is singled out because of its ubiquity and the fact that it represents more of an avoidance of interaction than interaction.

These different sums may be understood in terms of the game tree. The formula for $G + H$ shows that the game tree is built up in a rather complicated manner. Let $UG$ denote the set of all game positions for $G$. Likewise for $UH$. Then $U(G+H)=(UG)(UH)$. Next one adds edges as follows.  For any given game position $g$, let $T_{GL}(g)$ denote the subtree consisting of all subsequent left moves in $G$ with similar notation for right moves and for $H$. Then from any combined position $gh$ the set of subsequent edges is given as $T_{GL}(g)h \cup gT_{HL}(h)\cup T_{GR}(g)h\cup gT_{HR}(h)$. In the exclusive sum $G\hat  + H$, the game tree is given as the game tree of $G + H$ minus all edges and positions corresponding to  $G$ plays and $H$ plays appearing on the same board locations. The game tree $G \oplus H$ will be a proper subtree of the either the tree for $G+H$ or for $G \hat + H$.

In addition we have several different notions for the product of two games. The combinatorial product defined above involves non simultaneous play and corresponds to the arithmetic product when restricted to games that are also surreal numbers. In these additional products the key notion is that on any given play the player whose turn it is must play both $G$ and $H$ simultaneously. In the direct product $G \otimes H$ , $G$ and $H$ are played simultaneously but freely. In the exclusive direct product, $G \hat\otimes H$, $G$ and $H$ are played simultaneously but never on the same board site, which will in general necessitate some rule for breaking such moves. Finally there is a notion of an interactive product, $G \boxtimes H$ which again corresponds to simultaneous play but where moves are no longer free for each game but are restricted or even coupled to varying degrees giving rise to a collection of different games.

Sums would appear to best describe the generation of particles in superpositions of eigenstates since we want only a single informon to manifest at any step of game play. Products would better describe the situation of multiple particles since they allow multiple games to be played simultaneously, corresponding to the manifesting of multiple particles simultaneously. However there may be situations in which the generation of particles must occur sequentially and in such cases sums must be used. That might occur in the case of fermions since identical states are to be avoided. Since no such constraint applies to multiple bosons, they presumably may be described by products.

\subsection{Combinatorial Token Games}

The idea of a combinatorial game with tokens is used extensively so a few words are in order. Most combinatorial games are played on some kind of board using pieces or some kinds of marks which distinguish moves. Typical examples would be Chess, Checkers, Go, Hackenbush, Tic-Tac-Toe. A token is simply some kind of object that is placed on the board and which conveys information relevant to the play of the game. For example, chess pieces, by their association to specific roles, determine the kinds of moves available to them. Tokens may have mathematical or physical properties in their own right which can be useful to the play of the game. In specifying a combinatorial game with tokens we are considering situations in which tokens are either created or modified in the course of game play and the operations that may be performed upon these tokens enables one to construct combinations of games.  The reality games to be described below are played out on a causal manifold and tokens take the form of certain functions or vectors. Various operations may be defined on these tokens and used to define new games or to combine games. 
 
For example given a complex number $w$ and a token game $\mathbb{P}$ if we lay down a token $\phi$ in $\mathbb{P}$ then in the game $w\mathbb{P}$ we lay down a token $w(\phi)$, where some property of the token is modified by $w$. This enables the sums and products defined above to be expanded into more complex algebraic forms. In many cases, the number of tokens in a token game is constrained to some fixed value independent of the length of play. This is certainly true of a game like Chess. In other cases the number of tokens is determined solely by the possible length of play. In either case in a game $G$ formed as a sum of subgames $G_{i}$ according to $G=\sum_{i}w_{i}G_{i}$ one may interpret the term $w_{i}G_{i}$ to mean that the fraction of tokens assigned to the game $G_{i}$ is $|w_{i}|^{2}f(n)$ where $f(n)$ is the total number of tokens (as a function of length of game play $n$). In some situations such as the reality game played below, both interpretations may hold simultaneously, so that $w_{i}$ may modify both the number of tokens and some property of the tokens. 
 
\subsection{Ehrenfeucht-Fraiss\'e Games}

Games may beused to analyze the structure of mathematical theories and to compare structures within these theories. A brief digression to explore the idea of an Ehrenfeucht-Fraiss\'e will set the stage for the use of games to generate structures. The Ehrenfeucht-Fraiss\'e game \cite{hodges} appears in the study of mathematical logic where it is used to determine whether two structures may be viewed as expressing the same set of properties from the perspective of a specific logical theory. Mathematical logic consists of a collection of formal sentences constructed according to specified rules from an alphabet consisting of constant symbols, variable symbols, relational symbols, quantifiers and logical connectives. A sentence in formal logic has a counterpart in natural language but its formal nature makes it amenable to mathematical analysis. There are in addition a collection of rules which determine how one may create new sentences out of a pre-existing collection of sentences and which ensure that the new collection remains logically consistent and coherent. These are formal analogues of the laws of deduction taught in courses in philosophy and reasoning.

A first order language $L$ consists of a collection of symbols having different interpretations and formed into finite length strings according to a predetermined set of rules. The rules are designed to maintain consistency in the interpretation of these formulas or sentences. The basic symbols are constants $a,b,c, \dots$, variables $x,y,z, \dots,$ functions $F,G,H,\dots, $ relations $R,S,T, \ldots,$ =, and the logical quantifiers $\neg$, $\vee$, $\rightarrow$, $\leftrightarrow$, $\forall$, $\exists$. A term consists of a constant, variable, or function of constants and/or variables. A closed term has no variables. An atomic formula consists of s= t where s and t are terms or a relation of terms. A formula consists of a finite application of the logical quantifiers to a collection of atomic formulas. A variable is free if it is not within the scope of some quantifier. A sentence is a formula having no free variables. A theory is a collection of sentences. A model of a theory is a mathematical structure such that each constant in the theory corresponds to an element of the structure, each function and relation of the theory corresponds to a function and relation of the structure, and such that every sentence of the theory may be interpreted in the model and found to be true.

Often one wishes to understand the explanatory power of a theory. Does a theory, for example, describe everything about a particular model or are some features left unmentioned? Is the theory powerful enough to distinguish between specific models? An answer to the latter question can often be obtained through the play of an Ehrenfuecht-Fraiss\'e game.

Suppose that one is given two mathematical structures $A, B$, and one wishes to determine whether or not these two structures can be distinguished using a theory expressed in the language of first order logic. Assume that there are two players I, II and furthermore assume that play occurs for exactly n moves, where n is fixed in advance.

The game play is extraordinarily simple. Player I moves first and is free to choose any element they like from either $A$ or $B$. Player II then moves and may pick any element they like but only from the structure that Player I did not choose from. Play is repeated but with the caveat that at each step each player must choose a point that has not already been chosen. If there are no such elements to choose from then they simply forfeit their turn. Play continues in this way until a total of n steps have been played.

At the end of play one determines which of the two players has won the game. Let $a_{i}$  be the element of structure $A$ selected at the i-th move (whether by Player I or II) and let $b_{i}$ be the element of structure $B$ selected at the i-th move. One says that Player II wins the game if, whenever a relation R holds in $A$ for a sequence of elements $a_{i} ,a_{j} ,\ldots,a_{n}$  then it also holds for the corresponding elements $b_{i} ,b_{j} ,\ldots, b_{n}$  of $B$. Otherwise one says that Player I wins.

A strategy is a systematic procedure which tells a player how to move following a particular series of game plays. For example, one could simply pick an element at random. Usually one is interested in strategies that are deterministic, meaning that given a particular sequence of points selected in previous game plays there is a unique point to be selected on the current play. Such a strategy is called deterministic since the choices are fixed in advance. If Player II possesses a deterministic strategy which guarantees a win in n plays against Player I no matter how Player I plays, then we say that the game is determined and write $A \approx_{n} B$. 

Returning to theory, given a formal sentence $\phi$ and a model $A$, we write $A \vdash \phi$ if one can find elements, constants and relations in $A$ corresponding to those in $\phi$ so that the precise relations expressed by $\phi$ are satisfied by these corresponding relations in $A$. If for every logical formula $\phi$ having at most n quantifiers $A \vdash \phi$ if and only if $B \vdash \phi$, then we write $A \equiv_{n} B$.

The power of Ehrenfeucht-Fraisse games arises from the fact that $A \approx_{n} B$ if and only if $A \equiv_{n} B$. The game is often much easier to use to solve the logic problem than are logic tools alone. Thus games may be used heuristically without any ontological attribution being made as to the nature and status of the players. 

\subsection{Generative Games and Forcing}

Another important question facing logicians is to determine when a logical theory actually possesses a model and to exhibit such a structure. One of the most famous examples of this was the continuum hypothesis. This question concerns the sizes of sets and in particular whether the set of real numbers and the set of all subsets of natural numbers have the same size (cardinality). Cohen showed that is was possible to find models of set theory which extended the usual set theory, one of which satisfied the continuum hypothesis and one which did not. In this way he solved a long standing foundational problem in mathematical logic. The technique that he used to create such models involves a method called forcing \cite{hodges}.

The details are very technical but begin with the idea of a notion of consistency. A notion of consistency enables one to determine which theories actually possess models. Not all theories possess models. For example, the theory given by the single sentence $(a=b) \wedge \neg(a=b)$ has no model. Theories can be built up step by step provided that at each step one maintains consistency among the statements of the theory. This follows from the compactness theorem which states that if every finite subset of a theory $T$ possesses a model then the theory $T$ itself possesses a model. Building a theory step by step in this manner requires some notion of consistency. Formally, a notion of consistency $N$ is a collection of sets of sentences of $L$ which satisfy certain rules of logical consistency. For example, if $p \in N$ and $t$ is any closed term in $L$, then $p \cup \{t=t\}$ is in $N$. As an example involving a sentence, suppose that $\neg(\phi\wedge\sigma)$ lies in some subset $p \subset N$. Then either $p \cup \{\phi\}$ or $p \cup \{\sigma\}$ lies in $N$, but not both. There are seventeen such rules whose details are not necessary here (see \cite{hodges}). Each element $p \in N$ is called a condition. The idea is that each condition consists of a collection of formal sentences that are logically consistent. The important point is that if $N$ is a notion of consistency and $p$ is a condition of $N$, then $p$ has a model.

This is proven by virtue of a game. Assume that there are two players, I and II. The number of plays of the game is fixed in advance and described by some infinite ordinal number. The players alternate in making a move, Player I playing first. The goal of the game is to construct an increasing set of conditions possessing a model at each stage, and then forcing the final union of all of these conditions to have a model as well. At each stage of the construction different tasks are assigned according to each of the seventeen rules and these tasks are performed in such a way that only a finite number of new elements are added to the previously constructed condition. For example, one task might be as follows: given some condition $p$ constructed up to this point, one selects a closed term $t$ and, if it is not already present in $p$, one adds $t=t$ to $p$. Similarly, suppose that $p$ has already been constructed and that the formula $\neg(\phi\wedge\sigma)$ lies in $p$. Then this task might be to add either $\phi$ or $\sigma$, but not both. Whenever a limit ordinal is reached one simply assigns it the condition formed by taking the union over all previously constructed conditions. The tasks are each repeated a sufficient number of times to ensure that at the end of the construction no possible moves have been left undone. One possible strategy is to assume that a sufficient number of steps are carried out at each stage of the construction so as to parse at least once through the collection of all possible instances of all possible rules. Of course that will in general amount to a transfinite number of tasks to be performed at each stage of the construction and possibly a transfinite number of stages to complete the construction. That this procedure works is due to the recursive nature of the ordinals upon which this inductive process depends. Additional constraints may be placed on the choices made at each step. Finally some criterion is established which determines who wins the game. In other words the set of all possible sequences of play is partitioned into two disjoint subsets, one consisting of all wins for Player I and the other for all possible wins for Player II. 

One begins with a particular first order language $L$ and enlarges $L$ to form a new language $L(W)$ by adding a set $W$ of new constants, called witnesses. A notion of forcing for $L(W)$ is a notion of consistency $N$ which satisfies the following two conditions:
\begin{enumerate}
\item      if $p$ is a condition in $N$ and $t$ a closed term (meaning no variables) in $L(W)$ and $c$ is a witness which does not appear in either $p$ or $t$, then $p \cup \{t=c\}$ lies in $N$
\item      at most only finitely many witnesses appear in any $p \in N$
\end{enumerate}

Let us restrict ourselves to games in which there are only a countable number of steps. Let $P$ be some property that we would like our model to possess. One introduces witnesses and atomic formulae describing the expression of the property. We allow players I and II to alternate play as above, carrying out all of the necessary tasks and incorporating these witnessed formulae into the notion of consistency. If at the end of play the union of the chain of created conditions has property $P$ then we say that Player II wins. If Player II has a strategy which enables them to win no matter how Player I plays, then the property $P$ is said to be $N$-enforceable.

The importance of forcing is that it allows us to build up a structure step by step using a particular kind of game and ensure that it possesses a particular property.  The game approach is not only simpler in many cases than the axiomatic approach but it possesses the generative character that we seek for any model based on process theory. A more general technique for constructing classes of mathematical structures using games is presented in Hirsch and Hodgkinson \cite{Hirsch}.

\section{Archetypal Dynamics}
Although information has been considered to play a significant role in governing the behaviour of organisms for nearly a century, it is only recently that ideas of information began to appear in the physical literature based upon the connection between information and entropy as proposed by Shannon and Weaver \cite{Shannon}, which actually refers to the capacity for information and not its content. Ironically, information became a focus of interest only after it was stripped of any meaning. The program of archetypal dynamics is an attempt to provide a conceptual framework for studying the role of meaning laden information across disciplinary boundaries, particularly in those situations in which emergent phenomena appear. It postulates that the various entities of reality arise from and exist within particular conditions, that interactions among themselves and with the larger environment exhibit patterns, consistencies and constraints, all of which admit an effective conceptualization termed a semantic frame. The semantic frame gives meaning to the fundamental ontological questions of who, what, when, where, how and why. The semantic frame gives meaning to the entities and events and it is presumed that interactions between and with these entities are governed by flows of information whose meaning is imparted by the semantic frame.

In Archetypal Dynamics \cite{Sulis}, the behaviour of entities is held to be determined by salient, meaning laden information, which each entity detects and to which each entity responds according to its nature. The saliency of information is determined by each entity itself and is a consequence of its internal dynamics. Salience is considered to be a precursor to actual meaning. Salience is manifest in the phenomenon of transient induced global response synchronization (TIGoRS), in which an entity is capable of forming a differentiated pattern of responses in reaction to distinct patterns. Those patterns that induce the greatest convergence among responses are termed salient.  A related concept is that of compatibility, first introduced by Trofimova in her pioneering studies of emergent models of dynamical networks termed ensembles with variable structures (EVS), and used to determine when agents enter into the formation of dynamical linkages \cite{Trofimova1}. 

The Fundamental Triad of archetypal dynamics refers to realisations (the entities comprising the aspect of reality under consideration), interpretations (the semantic frames used by these entities or an observer to guide behaviours and interactions - exemplars of which are termed archetypes) and representations (the formal, linguistic or symbolic systems used to describe the realisation-interpretation relationship). 

Representations have generally taken the form of explanatory narratives, archetypal imagery, or mathematical models. The apparent success of many mathematical models, particularly in the physical sciences, has sometimes led to the belief that the mathematical depictions or descriptions of reality actually are reality.  This reification hides the fact that these mathematical models are idealizations of reality. From the standpoint of archetypal dynamics these mathematical theories constitute an interpretation of reality and particular mathematical models form archetypes. These archetypes are ideals to which reality may sometimes form a close approximation under particular conditions and circumstances. They are not reality. In the real world there is no infinity of entities, nor infinite volumes, temperatures, energies or masses. Nevertheless, under particular conditions some aspects of reality may behave in ways that closely mimic the idealization. Usually these conditions are those in which fluctuations of certain properties or the effects of extraneous influences or scales can be minimized for at least the duration of observations. Under such conditions one may think of reality as a finite approximation to the idealization (archetype) or conversely, the archetype as an infinite limit idealization of the reality.

Archetypal dynamics explicitly distinguishes between reality (realisation) and archetype (interpretation) and emphasizes the effective nature of the interpretation by taking notice of the particular conditions under which the semantic frame associated with the interpretation provides an effective description and interpretation of reality. Archetypal dynamics asserts that all physical theories, indeed all theories whatsoever, are at best effective theories, which hold under particular sets of conditions and interactions and demonstrate diminishing efficacy as these constraints are progressively violated. Even the most universal of physical laws come into question under conditions in which the symmetries underlying their existence fail to hold, or at extremes of scale where certain assumptions such as the continuity or existence of space-time come into question.

Rather than seeking a universal theory of everything, archetypal dynamics sees the world as being governed by a kaleidoscope of effective theories that interact with one another at the condition boundaries. Meaning becomes the relevant currency of exchange but meaning applies only conditionally and creativity arises in those regions where one set of conditions gives way to another. Emergence is viewed as a fundamental aspect of reality, with entities arising out of a co-creative interplay between realisation and interpretation, which serves to stabilize the conditions for their existence and persistence. 

In the archetypal dynamics perspective reality is always creative and in flux. Entities are conditional and therefore transient in nature. Entities, information, and meaning come into existence, persist, and fade away. There is a notion in the physical literature of a law of conservation of information but this is a bizarre idea derived from the unitary evolution of quantum systems and which certainly does not apply to meaning laden information. The idealizations that represent meaning can be thought of as existing in some Platonic universe but their applicability to reality fluctuates as the necessary conditions pass into and out of existence. The metaphysics of archetypal dynamics stands in contrast to the deterministic world view that has dominated the physical sciences. It posits a reality that is always in the process of becoming, that is always changing, that is fundamentally transient. 

\section{Interpolation Theory}

Interest in the use of interpolation theory was inspired by the work of Kempf \cite{Kempf} who used interpolation theory to provide a bridge between discrete and continuous representations of space-time and quantum fields. In physics, the state of a system is most commonly represented by a vector, defined over some field, usually the reals or complex numbers, together with an inner product, and having finite or countable components. These components are defined relative to a basis, which consists of a collection of vectors, none of which can be expressed as a sum of the others (independence). The number of vectors in the basis gives the dimension of the vector space. The components of a vector $\mathbf{v}$ are given by the inner product of $\mathbf{v}$ with the different vectors $\mathbf{v}_{i}$ that constitute the basis, i.e. $v_{i} = <\mathbf{v}, \mathbf{v}_{i} >$. The significance of the basis is that we can write each vector as a unique sum $\mathbf{v} = \sum_{i} v_{i} \mathbf{v}_{i} = \sum_{i} <\mathbf{v}, \mathbf{v}_{i} > \mathbf{v}_{i}$ . 

In classical mechanics each vector specifies a particular measurable property of the system, such as position or momentum.  In quantum mechanics each vector is infinite dimensional and is usually interpreted as giving a probability distribution from which the distribution of measurable properties may be obtained. 

In quantum mechanics the components $v_{i}$ of such a decomposition do not provide the results of any measurement directly. It is only the vector as a whole which can be ascribed a measurement outcome. Such vectors must therefore be treated holistically and do not represent any kind of generation or evolution.
 
The decomposition of vectors into a sum of basis vectors is a powerful technique mathematically which explains its widespread usage. But the notion of independence required to define a basis turns out to be too strong for many applications. This approach treats the vector space and basis vectors as having a prior existence and in the standard Fourier series approach the coefficients are defined as integrals over the base space of the basis functions. This may work in a static universe framework but certainly not in a process framework in which space-time is being generated. If one is to remain faithful to and consistent with the generative approach, then another method of generating functions is needed. Fortunately there is an alternative to the basis representation of vectors which is more general and which does permit a generative interpretation.

The simplest such approach is provided by the theory of function interpolation and in particular, by the Whittaker-Shannon-Kotel'nikov Theorem (WSK Theorem)\cite{Zayed}. The idea is to begin with a sampling of a function $f$ at a countable set of points $x_{k}$ and from the sampled values $f(x_{k})$ attempt to reconstruct the original function. The theorem originated in signal theory and versions are used today in the digitization and reconstruction of audio-visual signals. As it stands the idea is too vague since an infinite number of functions can be constructed from any countable set of values. In signal theory the normal way to limit this plethora is to restrict consideration to entire functions lying in $L_{2}(\mathbb{R})$ or $L_{2}(\mathbb{C})$ which are band limited, meaning that their Fourier transforms are non-zero within a bounded interval of frequency space, usually $[-\sigma,\sigma]$. The interpolated function is constructed as a sum over the sampled values of the form

\begin{displaymath}
f(x) = \sum_{k} f(x_{k})G_{k}(x,x_{k})
\end{displaymath}

Unlike the basis construction, the coefficients of this expansion do represent actual values of the function at the sampled points. In this representation one can think of the function as being constructed from a discrete set of 'events', namely the samples at the points $\{x_{1},\ldots, x_{k},\ldots\}$. Each event is interpreted as a function $f(x_{k})G_{k}(x,x_{k})$ and these are then superimposed to obtain the final function. The utility of this is that each actual occasion may now be interpreted as a continuous function which, unlike the occasion itself which is localized in space-time, now extends throughout the entire space-time. The functions of physics may now be seen as idealizations in which the number of samples is infinite and past, present and future information persists indefinitely. One may think of the function as being generated through the incorporation of ever more points into the construction and these points may be considered as temporal, spatial, or spatio-temporal as one likes. At first glance this might suggest that the past must persist into the future in order for these functions to be constructed but as they are merely interpretations of experience it is only necessary that information concerning the past persist in the present moment so that each current actual occasion is capable of generating an appropriate interpretation. Such information is already encoded in each actual occasion in the form of its content. Thus one may think of the function as being generated from the continual creation of actual occasions through the persistence of information residues.

The simplest interpolation models are those in which the functions $G_{k}(x,x_{k})$ are all derived from a single template function $G(x)$. The simplest means of doing so is through the use of a translation operator $T_{k}$ so that one may write 
\begin{displaymath}
G_{k}(x,x_{k}) = T_{k} G(x) = G(x-x_{k})
\end{displaymath}

One may then write the original equation in the form 

\begin{displaymath}
f(x) = \sum_{k}   f(x_{k})T_{k} G(x)
\end{displaymath}
where one may more easily consider $f$ to be constructed as a sum of primitive events each of the form $f(x_{k})T_{k} G(x)$. 

The Whittaker-Shannon-Kotel'nikov theorem asserts that if $f$ lies in $L_{2}(\mathbb{R})$ and its Fourier components lie within the range $0$ to $W$ for some frequency $W$ (band limited), and if we sample the function at a (infinite) set of discrete times $1/2W$ seconds apart, then we can write 

\begin{displaymath}
f(t) = \sum^{\infty}_{n=-\infty} f(n/2W) \sin\pi(2Wt-n)/\pi(2Wt-n)
\end{displaymath}

This series is in fact both absolutely and uniformly convergent on compact sets. This is a stronger result than for the usual Fourier series representations.

Writing $sinc(x) = \sin(x)/x$ we can rewrite the formula above in the following form:

\begin{multline*}
f(t) = \sum^{\infty}_{n=-\infty} f(n/2W) sinc(\pi(2Wt-n))=\\
= \sum^{\infty}_{n=-\infty} f(n/2W) T_{n/2W\pi} sinc(\pi(2Wt))
\end{multline*}

where $T_{n} f(x) = f(x-n)$.

Of course this will converge to the original function only in the case that one has a sampling over an infinite collection of points. In reality one will have only a finite sample to work with and so there will be errors arising from the undersampling. There are several results dealing with the effect of these truncation errors but the simplest is perhaps the earliest discovered. Sampling over $2N + 1$ points, we define the truncation error $T_{N} f(t)$ to be

\begin{displaymath}
T_{N} f(t) = f(t)-\sum^{N}_{n =-N} f(n/2W) \frac{\sin\pi(2Wt-n)}{\pi(2Wt-n)}
\end{displaymath}

Let $\sigma = 2W\pi$ , $-T \leq t \leq T$, $0 < \triangle t < (1/\sigma)$ and $E = \int^{\infty}_{–\infty} |F(\omega)|^{2} d\omega$ where $F(\omega)$ is the Fourier transform of $f(t)$. $E$ is called the total energy of $f$. Then

\begin{displaymath}
| T_{N} f(t)| = ((\sqrt{2})/\pi) E |sin(\pi t/\triangle t)|\sqrt{(T\triangle t)/(T^{2}- t^{2} ))}
\end{displaymath}

Observe that the truncation error decreases with both sampling frequency and sampling duration.

One may picture the original function as being a sum of wavelets, each being an oscillating function that goes to zero in the past and future and has a single global maximum. In the sum above, the maximum of each wavelet occurs at one of the discrete points given by $n$ and its value is given by the value of the function at that point.

This implies that the space of band limited functions, as a subset of the uncountable dimension vector space $\mathbb{R}^{\mathbb{R}}$ is actually isomorphic to the much smaller countable dimension vector space $\mathbb{R}^{\aleph_{0}}$ and approximately isomorphic to $\mathbb{R}^{n}$ for very large but finite $n$.

Any function that is band limited by $\sigma$ will be band limited by $\sigma^{\prime}$ for any $\sigma\leq\sigma^{\prime}$. It will thus have a representation in $\sigma'$ which will effectively over sample the function. This turns out not to be a problem as it simply introduces dependencies between the sampled values. It turns out that in such a case one may eliminate an arbitrary finite number of sampled values and still be able to reconstruct the original function \cite{Marks}.

One limitation of the method is that the requirement that $f$ be an entire function means that it cannot be simultaneously band limited and time limited. There is a form of uncertainty relation that holds in such a case even though signals in reality may be both time and band limited. This is a defect in mathematical modeling and again to serves to emphasize that these models are idealizations of reality and not exact models of reality.

We do have, however, the following theorem of Poussin: If $f$ is continuous and of bounded variation on $[a,b]$ and zero outside of this interval then

$$
f(t) = \lim_{n \rightarrow \infty} \sum_{t_{k}\in [a,b)} f(t_{k} ) \sin m(t- t_{k} )/m(t- t_{k} )
$$

\noindent where $t_{k} = k\pi/m$, $k=0, \pm 1, \pm 2,..$ and $m = n$ or $m = n+1/2, n=1,2,3,\ldots$…

The WSK theorem as originally stated refers only to functions of time. A multidimensional version of the WSK theorem was discovered by Parzen. If a function $f(t_{1} ,\ldots…,t_{N} )$ is band limited to a bounded N-dimensional rectangle with bounds $\sigma_{i}$ then one may write $f$ as:

\begin{widetext}
$$
f(t_{1} ,\ldots…,t_{N} ) = \sum^{\infty}_{k_{1} =-\infty}\cdots \sum^{\infty}_{k_{N}=-\infty} f(\pi k_{1} /\sigma_{1} ,\ldots…,\pi k_{N} /\sigma_{N} )sinc( \sigma_{1}t_{1} –\pi k_{1} )\cdots sinc(\sigma_{N}t_{N} –\pi k_{N} )
$$
\end{widetext}

A problem for the formulation of the WSK theorem is that time (and space) must be subdivided into a regular lattice whose spacings must relate to the frequencies of the sampling function. This suffices for an in-principle demonstration of this approach such as is being presented here but in reality there is no reason to expect that the actual occasions should come into being in precisely demarcated space-time intervals. It is more reasonable to assume the presence of at least a modest degree of variability in the relative positioning of occasions. This constitutes the case of non-uniform sampling. Extensive research has been devoted to studying non-uniform sampling, the effects of various kinds of information deficits, the use of different basis functions, the use of interpolation for solving various differential equations, the use of interpolation for wider classes of functions\cite{Zayed}. These topics are beyond the scope of this paper but are important for the development of this approach into a proper alternative to standard quantum mechanics. As an example of such an extension in the case that the function is separable there is a very strong result known as the Paley-Weiner-Parzen (PWP) Theorem.

Theorem: (PWP): Let $G_{i}$  be the entire function defined by

$$
G_{i}(t_{i}) = \negmedspace(t_{i}- t_{i,0}) \prod_{k_{i}} (1- t_{i}/ t_{i,k_{i}})(1- t_{i}/ t_{i,-k_{i}} ) \:i=1,2,..N
$$

Where $t_{i,k_{i}}$ are real numbers satisfying the estimate

$$
\sup_{k_{i}} | t_{i,k_{i}}- k_{i}\pi/\sigma_{i} | < \pi/4\sigma_{i} \; i=1,2,..N
$$

Then for any signal in separable form $f(t_{1},\ldots ,…t_{N} ) =  f_{i}(t_{1} )\cdots f_{N}(t_{N} )$ that is band limited to $\prod_{i} [-\sigma_{i} , \sigma_{i} ]$ we have

\begin{widetext}
$$
f(t_{1},\ldots ,…t_{N} ) = \sum_{1}\cdots \sum_{N}   f(t_{1,k_{1}},\ldots ,…t_{N,k_{N} })G_{1}(t_{1,k_{1}} )/ [G^{\prime}_{1}  (t_{1,k_{1}} ) (t_{1}, - t_{1,k_{1}})]\cdots G_{N}  (t_{N,k_{N}} )/ [G^{\prime}_{N}  (t_{N,k_{N}} ) (t_{N}, - t_{N,k_{N}})]
$$
\end{widetext}

Unfortunately no such explicit series is known in the case that f is non-separable.

\section{The Tapestry Formulation}

\subsection{Causal Tapestries}
The framework to be explored here for representing process is that of the causal tapestry. A causal tapestry is a discrete structure which represents the actual occasions that constitute the current transient now and process is represented as a combinatorial forcing game with tokens (reality game) which generates the next causal tapestry, corresponding to the fading of the current now and the creation of the next. The formal structure of the causal tapestry has been described extensively elsewhere \cite{s7} and presented in more detail in the appendix. The causal tapestry draws upon ideas concerning the role of past information in determining the present \cite{sorkin,Fotini, Dowker} so as to model a relativistically compatible transient now. The basic element of a causal tapestry is the informon, which takes the form $[n]<\alpha>\{G\}$ where $n$ is a descriptive label (representing the realisation of actual occasions), $\alpha$ is an ordered tuple (representing the interpretation) having the general form $(\mathbf{x},\phi,p,f)$ where $\mathbf{x}$ is an element of a causal manifold, $\phi$ an element of a Hilbert space, $p$ a collection of properties, for example lepton number, charm and the like, and $f$ is an element of a Lie algebra, and $G$ is a causally ordered set of \textquoteleft prior\textquoteright$\:$ infomons. The causal set $G$ represents all of the information entailed within those prior informons (actual occasions) that materially contribute to the creation and existence of the current informon (actual occasion). Interpreted, it will, in general, be a subset of the past causal cone of the embedded informon. The causal tapestry embeds into the causal manifold as a space-like hyper-surface and the elements $\phi$ form a collection of interpolation samples of a wave function. 

In order to properly describe reality from a process theory perspective it is necessary to consider separately the space of actual occasions and the processes that generate them. There are two fundamental causal tapestries needed to describe a physical process - an event tapestry, which generalizes the idea of a position space, and a transition tapestry, which generalizes the idea of momentum space. These two tapestries are linked through duality relations to form a reality tableau. The formal details are presented in the appendix. Processes are implemented by means of the reality game and therefore each process may be considered as a game in its own right. The global the reality game arises from the interactions among all of these individual process games. Processes generate actual occasions but processes themselves may be rendered active or inactive through the arising and fading away of configurations of actual occasions. The processes will thus form an ordered structure reflecting game play and the activation or inactivation of these processes.

The reality game is based upon the idea of forcing. In a reality game a pair of players successively carry out a set of tasks resulting in the creation of informons and interpretation maps while remaining consistent with some particular set of dynamical constraints, usually given by means of some group or semigroup of symmetries such as a differential equation. 

The reality game generates a succession of reality tableaux. Let $(\Omega,\Pi)$ denote a reality tableau. That is, $\Omega$ is an event tapestry $(L, K, M \times F(M) \times D \times P(M^{\prime}), I_{p} )$ and $\Pi$ is a process tapestry $(L^{\prime}, K^{\prime}, M^{\prime} \times F(M^{\prime}) \times D^{\prime} \times P^{\prime}(M), I^{\prime}_{p})$ for which the tableau duality postulate and the tableau consistency criterion hold. By a reality game is meant a game which is given $\Omega$ and $\Pi$ and at the end of play has created new tapestries $\Omega^{\prime}$ and $\Pi^{\prime}$ where $\Omega^{\prime} = (L^{\prime\prime}, K^{\prime\prime}, M \times F(M) \times D \times P(M^{\prime}), \Omega \cup I_{p})$ and $\Pi^{\prime} = (L^{\prime\prime\prime}, K^{\prime\prime\prime}, M^{\prime} \times F(M^{\prime}) \times D^{\prime} \times P^{\prime}(M), \Pi \cup I^{\prime}_{p})$ such that $(\Omega^{\prime},\Pi^{\prime})$ form a reality tableau. Reality tableaux are defined recursively and so require the assumption of some prior set of prior tableaux upon which the reality game is to be played. This prior set may be empty or may be any self consistent set of tableaux. 

The effect of the game is to create new sets of informons and new sets of prior tapestries but with the interpretation structures left intact. The interpretation structures remain invariant because, as discussed in the introduction, they provide the archetypal frame of meaning underlying all information flow in the system. In order that these frames of meaning remain consistent and coherent through the history of the system it is necessary that they remain invariant under the play of the game. A change in the interpretation structures is interpreted as a fundamental change in some aspect of the system that results in a significant difference in behaviour and properties, for example a change of phase, or a change of symmetry or a change of distinguishable constituents. It is true that in general, information may evolve in a such a manner as to force the emerge of a new semantic frame and thus new interpretation structures. As noted above, the analysis of such transitional situations is beyond the scope of the current work. 

\section{Non Relativistic Quantum Mechanics}

In this section we consider a simplified model of NRQM in (3+1) dimensions. While technically unsophisticated it nevertheless provides an in principle demonstration of this approach. We utilize sinc interpolation to generate the state space interpretation and we do so on a regular lattice. In general one would expect that a nonuniform lattice would be more realistic, but in the case of sinc interpolation it turns out that even if nonuniform sampling is used, the sample values can be transferred to a nearby uniform lattice and still achieve a good approximation \cite{Maymon} so for ease of construction one might as well utilize a regular lattice from the start. In NRQM the causal structure is quite simple, being the ordered product of $\mathbb{R}^{3}$, treated as a antichain, and $\mathbb{R}$, the time component which generates the order. The geometry is flat, so that the causal manifold is just $\mathcal{M}=\mathbb{R} \otimes \mathbb{R}^{3}$. The state space is just the usual Hilbert space  $\mathcal{H}(\mathcal{M})$ of complex single valued $L^{2}$ functions on $\mathcal{M}$. The lattice decomposition of $\mathcal{M}$ consists of all points of the form $(n_{0}t_{P},n_{1}l_{P},n_{2}l_{P},n_{3}l_{P})$ where $t_{P}$ is the Planck time and $l_{P}$ the Planck length and the $n_{i}$ are integers.

We will embed each causal tapestry into a space-like slice of the form $(t,\mathbf{n})$ for fixed $t$ and $\mathbf{n}\in \mathbb{R}^{3}$. Thus each causal tapestry corresponds to a space-like hypersurface of $\mathcal{M}$. Previous work has shown that any game played out on causal tapestries that respects the causal structure will also respect Lorentz invariance \cite{s7}. Thus each causal tapestry represents a \textquoteleft transient now\textquoteright$\:$ with respect to the causal structure. This is not equivalent to a notion of simultaneity in the coordinate structure; that is not preserved under Lorentz transformations.

In order to derive a connection with NRQM we utilize the space-time approach developed by Feynman and expressed in his path integral formulation of quantum mechanics. There are two main reasons for this. The first and most important is that the causal tapestry approach provides an emergent generation of space-time and generates state space functions on a space-time point by point basis and thus a version of quantum mechanics which provides information at specific space-time points facilitates a matching between these two approaches. Secondly, the Lagrangian approach is, as Feynman pointed out, more general than the Hamiltonian approach and so has wider applicability. Several authors have pointed out how the wave function can be constructed based upon past data \cite{Mann,sorkin,Fotini,Dowker}. 

Let us therefore consider the motion of a single particle under a Lagrangian $L$. Denote the reality game for the particle as $\mathbb{P}$. Since we are dealing with a single particle we shall simplify the discussion and focus just on the event tapestry. Assume that a prior causal tapestry $\mathcal{I}_{p}$ has been generated with current tapestry $\mathcal{I}$. We consider the next round of game play. The game is played for $n$ plays. The goal at the end of game play is to have constructed a new causal tapestry, $\mathcal{I}'$ having informons of the form $[n]<\alpha>\{G\}$ where $\alpha=(\mathbf{m},\phi,\mathbf{p},\tau)$ and $G$ is an ordered set derived from $\mathcal{I}_{p}\cup \mathcal{I}$ and $\phi$ provides a interpolation component of the wave function of the particle sampled at the point $\mathbf{m}$. Since we are concerned solely with the event tapestry we shall ignore the $\tau$ component. We shall also consider the particle as being in a single eigenstate so that we may ignore the $\mathbf{p}$ component as well. The more general case will be discussed below. 

Player 1 is free to choose any prior informon, say $[n]<\alpha>\{G\}=[n]<((t_{n},\mathbf{x}),\phi_{n})>\{G\}$ where $(t_{n},\mathbf{x})$ is the lattice embedding, and $\phi_{n}$ is the partial wave function contribution. The partial wave function has the form $\Theta_{n}(\alpha)sinc(\alpha)$ where $\Theta_{n}(\alpha)$ is a constant dependent upon the embedding point of $\alpha$ and $sinc(\alpha)$ is shorthand for the product of sinc functions centered around the embedding point of $\alpha$. Player 2 selects a point on the next lattice layer $(t_{n}+t_{P}, \mathbf{y})$ where $\mathbf{y}$ is any $\mathbb{R}^{3}$ point such that $(t_{n}+t_{P}, \mathbf{y})$ lies on the lattice. The only strategy that Player 2 requires is to ensure that the pairing of prior and current points has not been previously chosen.  Player 1 now constructs $\phi_{m}$. The strategy for Player 1 is as follows. 

Given the Lagrangian $L$, Player 1 calculates the action corresponding to a straight line path from the point $(t_{n},\mathbf{x})$ to the point $(t_{n}+t_{P},\mathbf{y})$. Let $\beta =((t_{n}+t_{P}, \mathbf{y}),)$ be a partial interpretation. Denote this integral as $S[\beta,\alpha]= \int^{(t_{n}+t_{P},\mathbf{y})}_{(t_{n},\mathbf{x})} dt L$. Now place a token of the form $(t_{p}l^{6}/A^{4})e^{iS[\beta,\alpha]/\hslash}\Theta_{t_{1}}(\alpha)sinc(\beta)$ on the site $(t_{n}+t_{P}, \mathbf{y})$. $A$ is a Feynman-Hibbs path integral normalization constant. There may be existing tokens on the site. If so, simply add it to the collection.

Finally Player 2 must determine a new content. Add a second form of token consisting of an ordered sum of the form $\hat G + \{[n]<\alpha>\{G\}\}$ where $\hat G$ is some up-set of $G$. Again if there are pre-existing such tokens, then add it to the collection.

Play continues until the allotted number of allowed plays is reached. Note that a player does not have to make a move on any given play if no appropriate move is available. At the end of game play a new informon is created which embeds to this site. First any index element not previously used is selected, say $k$. Then the interpretation $\beta$ is completed, having the form $\beta = ((t_{n}+t_{P},\mathbf{y}), \phi)$ where the state function $\phi = \sum_{\alpha}(t_{p}l^{6}_{p}/A^{4})e^{iS[\beta,\alpha]/\hslash}\Theta_{t_{1}}(\alpha)sinc(\beta)$ with the sum being taken over all state space tokens on the site and $A$ is a normalization factor described by Feynman and Hibbs \cite{Feynman}. Finally a content $H$ set is chosen having the form $H= \cup \{\hat G + \{[n]<\alpha>\{G\}\}\}$ where the union is taken over all content tokens. The new informon is thus $[k]<\beta>\{H\}$. This is done for every point on the new lattice slice on which there are tokens. This collection of informons $\mathcal{I}'$ becomes the new current causal tapestry and the new prior tapestry set becomes $\mathcal{I}\cup \mathcal{I}_{p}$.

Let $L_{1}$ denote the collection of lattice sites corresponding to the embeddings of informons of the prior tapestry $\mathcal{I}$. Feynman and Hibbs \cite{Feynman} show that, if $\Psi_{t_{1}}$ and $\Psi_{t_{2}}$ are wave functions for a particle at times $t_{1}< t_{2}$ respectively, then $\Psi_{t_{2}}$ can be calculated from $\Psi_{t_{1}}$ using the transition kernel as follows:

\begin{displaymath}
\Psi_{t_{2}}(x)=\int K(x,t_{2}:y,t_{1})\Psi_{t_{1}}(y) dV
\end{displaymath}

In the case where $|t_{2}-t_{1}|=t_{P}$ corresponding to the time difference between two successive lattice slices (causal tapestries), and $dV=l^{3}_{p}$, the spatial distance between adjacent lattice sites, the above integral can be approximated on the lattice slice as 

\begin{multline*}
\Psi_{t_{2}}(x)\approx \sum_{y\in L_{1}} K(x,t_{2}:y,t_{1})\Psi_{t_{1}}(y) l^{3}_{p}=\\
\sum_{y\in L_{1}} K(x,t_{2}:y,t_{1})\sum_{y'\in L_{1}}\Theta_{t_{1}}(y')sinc(y')(y)l^{3}_{P}=\\
\sum_{y\in L_{1}} K(x,t_{2}:y,t_{1})\Theta_{t_{1}}(y)l^{3}_{p}=\\
\sum_{y\in L_{1}} \int e^{\frac{i}{\hslash}S[(x,t_{2}),(y,t_{1})]}\mathcal{D}y\Theta_{t_{1}}(y)l^{3}_{p}\approx\\
\sum_{y\in L_{1}}e^{\frac{i}{\hslash}S[(x,t_{2}),(y,t_{1})]} \Theta_{t_{1}}(y)t_{p}l^{6}_{p}/A^{4}
\end{multline*}

\noindent where in the last line above the action $S[(x,t_{2}),(y,t_{1})]$ is calculated on a straightline segment between $(x,t_{2})$ and $(y,t_{1})$ and $K(x,t_{2}:y,t_{1})= \int e^{\frac{i}{\hslash}S[(x,t_{2}),(y,t_{1})]}\mathcal{D}y\approx
e^{\frac{i}{\hslash}S[(x,t_{2}),(y,t_{1})]}t_{p}l^{3}/A^{4}$ using the Feynman and Hibbs approximation to the path integral (again S calculated on a straightline path). Utilizing Parzen's theorem \cite{Zayed}, the wave function may be approximated on the full spacetime slice $\{t_{2}\}\times \mathbb{R}^{3}$ as $\Psi_{t_{2}}(z)\approx \sum_{y\in L_{2}} \Psi_{t_{2}}(y)sinc(y)(z)=\sum_{y\in L_{2}}\!\sum_{y'\in L_{1}}e^{\frac{i}{\hslash}S[(y,t_{2}),(y',t_{1})]} \Theta_{t_{1}}(y')t_{p}l^{6}_{p}/A^{4}sinc(y)(z).$

In causal tapestry terms this becomes 

\begin{displaymath}
\Psi_{t_{2}}(z)\approx \sum_{\beta\in \mathcal{I}'}\![\sum_{\alpha\in \mathcal{I}}e^{\frac{i}{\hslash}S[\beta,\alpha]} \Theta_{t_{1}}(\alpha)t_{p}l^{6}_{p}/A^{4}]sinc(\beta)(z).
\end{displaymath}

In other words, the sum of the state space elements across the causal tapestry approximates the NRQM wave function on the spacetime slice in which the causal tapestry embeds. Moreover, Parzen's theorem also shows that in the limit of perfect transfer of information (infinite game play) $(n\rightarrow \infty)$ and infinitesimal sampling of space-time $(l_{P},t_{P}\rightarrow 0)$, this sum converges uniformly and absolutely to the wave function on the lattice slice.  This shows that the reality game model provides a discrete approximation to the path integral and converges to the proper path integral asymptotically, showing in turn that NRQM emerges in the causal tapestry approach in the asymptotic limit. Thus the reality game provides a more fundamental theory than does NRQM, which becomes an emergent and effective theory. Even far from the asymptotic limit, the causal tapestry approach can yield highly accurate approximations to NRQM. 

In the case that the particle has energy and momentum bounded well away from the Planck limits, the difference in values between the causal tapestry can be immeasurably small. For example, at the sampling frequencies considered here, namely Planck scale, using the Yao and Thomas theorem \cite[pgs 91-92]{Zayed}, the truncation error is approximately $\frac{t_{P}l^{3}_{P}}{\pi^{4}TL^{3}}\approx 2 \times 10^{-151}/TL^{3}$, where $t_{P},l_{P}$ are the numerical values of the Planck time and Planck length respectively and $T,L$ are the durations and spatial extensions of the observations. Thus for an observation at the smallest spatial and temporal intervals that can currently be measured, this would give an error on the order of $2 \times 10^{-104}$ which renders the Schr\"odinger wave function and the causal tapestry wave functions effectively identical.

This also shows that we need not assume that the sampling occurs at Planck scales and may treat the temporal and spatial generating frequencies as independent variables and adjust the scales upwards until a significant discrepancy occurs. For example, accepting a discrepancy of say $10^{-10}$ this could allow for generating intervals of up to $10^{30}t_{P}$ seconds and $10^{20}l_{P}$ meters.

Note also that in this model of NRQM the only information required to determine the informons of the new causal tapestry are the informons of the current causal tapestry, so that the content of each informons need only consist of informons from the preceding causal tapestry. Thus in NRQM the system evolves from one causal slice to the next, in other words from one transient now to the next. The situation is more subtle in the case of relativistic quantum mechanics where the content set must come (nontrivially) from the past light cone of each informon and the embedding into the causal manifold takes a more complicated form.

\subsection{Superposition States: The Case of Multiple Subprocesses}

The scenario above applies to the case of a simple process for a particle in an eigenstate. In the case of a particle in a superposition of eigenstates the play of the game is more subtle. Suppose that the wave function of the particle $\Psi = \sum_{i} w_{i}\Psi_{i}$ where $\Psi_{i}$ are eigenfunctions of some operator and $w_{i}$ are the usual weights.

The process generating the particle can be considered as a sum of subprocesses, each generating an eignestate. Then we may write the process as $\mathbb{P} = \hat \sum_{i} w_{i}\mathbb{P}_{i}$ where the $\mathbb{P}_{i}$ are the corresponding subprocesses and the weights $w_{i}$ are interpreted as for sums of token games. Note that the exclusive sum is used here. This ensures that an actual occasion corresponding to a single eigenstate, meaning a single ontological state, of the particle manifests at any particular lattice site. This means that the particle itself is never in a superposition of ontolgically distinct states; it is only the process that generates the particle that can be in a superposition. In order to keep track of which subprocess is manifesting in a particular informon we must now include properties in the interpretation of each informons. Here it suffices to add a property $i$ which means that the $i$-th subprocess applies to this informon. The interpretation of an informon will now take the form $\alpha=(\mathbf{m},\phi,i)$. Each site will therefore also acquire a token for the process, here $i$.  Except when a new informon is generated, a subprocess may only be played on sites on which the token corresponds to that subprocess.

Play proceeds more or less as in the previous case. On any given play one of the subprocesses is chosen to be played, say $\mathbb{P}_{i}$. When Player 1 makes their first move they may only extend from an informon whose interpretation is of the form $\alpha=(\mathbf{m},\phi,i)$. If they extend to a token naive site, the site is given a token $i$. If the site already contains tokens then it must contain only $i$ tokens. This is important when relativistic conditions arise because of the more complicated structure of the causal space. This emphasizes the role of the content set which carries relevant causal information from the causal past into the present. The idea is that an extension of a process from the causal past into the future involves a transition from informons previously generated by the process to informons currently being generated by the process. This process is somewhat akin to a percolation front except that the updating of the front occurs sequentially and in a saltatory manner. Informons thus propagate outwards from informons of origin.

Once the above consistency check has taken place the remainder of play is as described above since effectively now only a single subprocess is involved.

The selection of sites for the individual subprocesses may be random or in the case of only a finite superposition of subprocesses (say n) the lattice may be subdivided into multiple regular lattices but with a much larger lattice separation, ($nl_{P}$). When one moves to the emergent level of the global wave function one notes the following. Due to the consistency criterion, the interpolated wave function on the content of a given informon having property $i$ will correspond to an approximation of the eigenstate wave function $\Phi_{i}$ on the past causal cone of its embedding point. On the current causal tapestry $\mathcal{I}$, the interpolation has the form 

\begin{multline*}
\Phi_{\mathcal{I}}=\sum_{[n]<\alpha>\{G\}\in \mathcal{I}} w_{i}\Theta_{n}sinc(\alpha)=\\
\sum_{i}\sum_{[n]<\alpha>\{G\}\in \mathcal{I}|\alpha=(\mathbf{m},\phi,i)} w_{i}\Theta_{n}sinc(\alpha)\approx\\
\sum_{i} w_{i}\Phi_{i}\approx \Phi
\end{multline*} 

Thus the interpolation subdivides into a sum of interpolations defined over disjoint sublattices of the current lattice slice. This is also true for the global interpolation taken over the entire history of causal tapestries. On each of these sublattices the interpolation provides an approximation to one of the weighted eigenfunctions. Each of these interpolations may provide an accurate approximation to the actual eigenfunction over the causal manifold and thus their sum to the wave function itself. In the case of a random distribution of lattice sites supporting the interpolation one takes note of the fact that for wave functions corresponding to particles having energy and momentum bounded far away from the Planck scales, these interpolations correspond to a rather significant oversampling of the individual eigenfunctions and one may use a result of Marks \cite{Marks} which shows that in such a case the missing samples may be reconstructed from the given samples, so that there is in effect no missing information about the eigenfunction and it may be reconstructed perfectly. In the case of the regular lattice decomposition one notes that for a superposition of a reasonably small number of eigenfunctions, the lattice spacing will still provide a sampling of the original function at a lower frequency, but so long as this frequency still exceeds the Nyquist frequency for the eigenfunction, it will provide an accurate interpolation.

Thus at the level of the informons, when the causal tapestry is generated by the reality game at every step of play only a single informon is being generated and it will correspond to a single well defined ontological state of the particle. As play proceeds the particle will manifest all of the eigenstates individually and distributed throughout the causal tapestry, and so effectively throughout the causal manifold. At the level of the observer one will observe only the interpolated wave function, which will correspond to a superposition of these eigenstates. There is no paradox here. The problem arises as a result of the inherent inability to resolve reality at the level of the individual actual occasion. One can have a realist dynamic operating at the lowest level even though at the level at which we can observe reality one sees only these seemingly paradoxical superpositions. The superpositions reflect the nature of the generating processes, not the nature of the reality being generated.

\section{Reality Game Model of Measurement}

Bohr understood the process of measurement to provide a correspondence between some aspect of a quantum system and the behaviour of a classical measurement apparatus. This conforms to the actual situation when an experimenter carries out a measurement. Not every apparatus can serve as a measurement apparatus. For example, one could place a dial on one bob of a kicked double pendulum and attempt to measure the momentum of a particle by letting it collide with the kick plate. The subsequent motion of the dial will be chaotic, so that even if a particle with exactly the same momentum hits the plate, the resulting motion of the dial will be effectively random and decidedly non-Gaussian, so no single \emph{measurement} value can be determined. Rather a measurement apparatus must have the property that an activation (or interaction) with the device results in a transition of states within the apparatus that ultimately results in a single stable fixed state, from which can be derived some token like a mark or a dial position which can then be translated into a measurement value. Dynamically this means that the phase space of the measurement apparatus must subdivide into multiple stable basins of attraction all leading to fixed points asymptotically and every activation or interaction will place the apparatus into one of these basins of attraction. Moreover, there must be a stable correspondence between the activation and the subsequent basin, so that identically prepared activations will result in the apparatus entering into, more or less, identical basins of attraction. In the quantum mechanical setting this means that if copies of the quantum system can be placed into identical eigenstates then the measurement apparatus will be in identical basin of attractions after interaction. The basins should be stable, attracting, and separated. In particular they should not be tangled or lead to chaotic behaviour. The size of the basin determines the resolution of the apparatus. Once a triggering or interaction has taken place the apparatus should evolve within the selected basin until the fixed point has been reached. In other words the potential well associated to the basin should be sufficiently deep so as to prevent the apparatus from jumping from one basin to another, even in circumstances in which the triggering entity continues to interact with the apparatus over an extended period of time.

These are general considerations which appear reasonable to demand of any apparatus purporting to be a measurement apparatus. In the model of measurement outlined below we model only these general features and not the specifics of any particular apparatus. 

Let us consider the situation of a single particle interacting with a measurement apparatus. The particle and the apparatus are modeled by distinct reality games, $\mathbb{P}$ and $\mathbb{M}$ respectively.  We assume that the particle is in a superposition of eigenstates $\Psi_{i}$ (with eigenvalues $\lambda_{i}$) of its Hamiltonian. In NRQM the state of the particle would be represented as $\Psi = \sum_{i}w_{i}\Psi_{i}$ for some weights $w_{i}$. We assign a distinct game $\mathbb{P}_{i}(\lambda_{i})$ to generate each eigenstate and consider the particle game as a weighted combinatorial sum of these lesser games, $\mathbb{P} = \sum_{i} w_{i}\mathbb{P}_{i}(\lambda_{i})$. The weight in this case modifies the value of the state interpretation of each informon generated. Likewise treat each basin of attraction of the measurement apparatus as a distinct subsystem and assign to each a distinct game which generates its own dynamic. If the set of possible measurement values is $\{\mu_{j}\}$,
then to each value we assign a game $\mathbb{M}_{j}(\mu_{j})$ and we consider $\mathbb{M} = \sum_{j} \mathbb{M}_{j}(\mu_{j})$ where the appropriate normalization is subsumed within $\mathbb{M}_{j}(\mu_{j})$. 

The fundamental algebraic structure of functional analysis is that of a vector space, and just as a vector in a finite dimensional vector space may be decomposed as a sum of basis elements, so can any element of a Hilbert space. Generally a Hilbert space will possess multiple distinct bases and so given two such bases $\{\phi_{i}\}$ and $\{\sigma_{i}\}$ we may write any element $\Psi$ as $\Psi=\sum_{i}w_{i}\phi_{i}=\sum_{j}v_{j}\sigma_{j}$. In functional analysis one is free to choose any basis at will and represent any vector within that basis. In the ideal world of mathematics this occurs without any physical implications. Again the functional analytic framework captures an algebraic relationship but without any aspect of ontology. The mathematics demonstrates that which is implicit in the structure of the wave function and what is to be expected should such a change in measurement basis be carried out but it doesn't specify how this change is to be realized in reality. While a change of basis in a finite dimensional vector space can be viewed merely as a change of reference frame, or perspective, a change of basis in quantum mechanics represents a change of measurement serving as the underlying basis for the observation. It does not merely represent a change of observer or of position of an observer. It reflects, rather, a fundamental change in how the observation is being carried out and what kind of measurement apparatus is being utilized to carry out the observation. 

In the reality game perspective, a decomposition of a wave function as a sum of the form $\Psi=\sum_{i}w_{i}\phi_{i}$ implies that the actual occasions that underly the particle are being generated by a process which can be decomposed into a sum of simpler processes as $\mathbb{P} = \sum_{i} w_{i}\mathbb{P}_{i}$ where each subprocess $w_{i}\mathbb{P}_{i}$ generates the wave function contribution $w_{i}\phi_{i}$. A change of basis therefore implies not merely a change of perspective but a change in the very processes that are generating the actual occasions. That is, a change of basis carries ontological significance and is not simply epistemological.

Let us fix a context within which to observe a quantum system and let that system experience an evolution described by a wave function $\Psi$. Furthermore, let there be two distinct basis decompositions of $\Psi$, namely, $\Psi = \sum_{i}w_{i}\Psi_{i}=\sum_{j}w'_{j}\Psi'_{j}$.  There will be distinct reality games (processes) $\mathbb{P}_{i}$ generating $\Psi_{i}$ and $\mathbb{P}'_{j}$ generating $\Psi'_{j}$. Thus there will be corresponding reality games $\mathbb{P} = \sum_{i}w_{i}\mathbb{P}_{i}$ and $\mathbb{P}' = \sum_{j}w'_{j}\mathbb{P}'_{j}$, both of which generate $\Psi$. We say that the reality games (processes) $\mathbb{P},\mathbb{P}'$ are $\Psi$-equivalent and write $\mathbb{P}\equiv_{\Psi}\mathbb{P}'$. Note that under a different context it may be the case that these reality games no longer generate the same wave function. This is a result of the non-commutativity of many quantum mechanical operators. 

For simplicity let us ignore details of experimental error. The measurement process is understood as taking place in three stages. In the first stage, prior to the initiation of any interaction between the particle and the apparatus, the two games are played freely, so that the combined process can be represented as $\mathbb{P}\otimes \mathbb{M}=(\sum_{i}w_{i}\mathbb{P}_{i})\otimes \mathbb{M} = (\sum_{i}w_{i}\mathbb{P}_{i})\otimes (\sum_{j}\mathbb{M}_{j}(\mu_{j}))=\sum_{i}\sum_{j}w_{i}\mathbb{P}_{i}\otimes \mathbb{M}_{j}(\mu_{j})$. 

In the second stage, the particles enters into the region of the measurement apparatus. In this stage there is no exchange of physical attributes such as energy, momentum, spin, mass etc. Instead the measurement apparatus establishes a new set of boundary conditions which results in a purely informational effect upon the processes generating the particle. This information results in a reversible change in the processes generating the wave function. This occurs, as described above, because it is precisely those particle states that are eigenstates of the operator corresponding to the measurement apparatus that are preserved under an interaction with the apparatus. Only these states can survive an interaction with the apparatus for a long enough period of time so that a measurement can take place.  Although the processes generating the wave function have changed, the observed wave function remains unchanged. We may now write this new process as $\mathfrak{M}(\mathbb{P})=\mathbb{P'} =\sum_{j} w'_{j}\mathbb{P'}_{j}(\lambda_{j})$ where $\mathbb{P'}_{j}$ represents the process that generates the eigenfunction $\Psi_{j}(\lambda_{j})$ corresponding to the eigenvalue $\lambda_{j}$ of the measurement apparatus. Here $\mathfrak{M}(\mathbb{P})$ denotes that the process $\mathbb{P}'$ was derived from $\mathbb{P}$ under a change of basis to that of eigenfunctions of the measurement apparatus. Since the wave functions generated by $\mathbb{P}$ and $\mathfrak{M}(\mathbb{P})$ are related by a change of basis, they represent the very same wave function, $\Psi$ and so $\mathbb{P}\equiv_{\Psi}\mathfrak{M}(\mathbb{P})$.  Although this interaction is purely information, it nevertheless results in a transition of processes, which we write as:

$\mathbb{P}\otimes \mathbb{M}=\sum_{i}\sum_{j}w_{i}\mathbb{P}_{i}\otimes \mathbb{M}_{j}(\mu_{j})\rightarrow \mathfrak{M}(\mathbb{P})\otimes \mathbb{M}=\sum_{k}\sum_{j}w'_{k}\mathbb{P}'_{k}(\lambda_{k})\otimes \mathbb{M}_{j}(\mu_{j})$.

\noindent 

Now each subprocess of $\mathfrak{M}(\mathbb{P})$ can couple only with certain subprocesses of the measurement apparatus (corresponding to nearby measurement values) so that the combined process at the initiation of measurement will be represented more accurately as  $\mathfrak{M}(\mathbb{P})\boxtimes \mathbb{M} = \sum_{j} \sum_{j'\in H(j)}(\tilde w_{jj'}\mathbb{P}'(\lambda_{j})\otimes\mathbb{M}_{j'}(\mu_{j'}))$ where $H(j)$ is the set of measurement apparatus states to which the particle state $i$ couples. In the case of no error this becomes $\mathbb{P}\boxtimes \mathbb{M} = \sum_{j}\tilde w_{j}\mathbb{P}'(\lambda_{j})\otimes \mathbb{M}_{j}(\lambda_{j})$.

In the third stage the particle enters into the region in which a physical interaction with the measurement apparatus becomes possible. The actual initiation of interaction becomes a function of the coupling between particle and apparatus. That is, with each play of the game there will manifest a particle informon, say $[n]<\alpha>\{G\}$. Let $\alpha = (\mathbf{m},\phi,\mathbf{p})$. By assumption, the state function $\phi$ of this informon will be generated by some process of the form $\tilde w_{jj'}\mathbb{P}'(\lambda_{j})$ for some eigenvalue $\lambda_{j}$, so that it will be an interpolation contribution to $\tilde w_{jj'}\Psi'_{j}(\lambda_{j})$, the eigenfunction corresponding to $\lambda_{j}$. Note that $\phi(\mathbf{m})=\tilde w_{jj'}\Psi'_{j}(\lambda_{j})(\mathbf{m})$. There will also manifest a measurement apparatus informon, say $[m]<\beta>\{H\}$ which is being generated by some subprocess, say $\mathbb{M}_{j'}(\mu_{j'})$. The likelihood that the particle process will couple to the measurement apparatus process will be expected to depend upon the value of the interpolation component at that site and so should be proportional to $\phi^{*}(\mathbf{m})\phi(\mathbf{m})=\tilde w^{*}_{jj'}\Psi'^{*}(\lambda_{j})(\mathbf{m})w_{jj'}\Psi'(\lambda_{j})(\mathbf{m})$.
The likelihood that it will fail to couple is thus proportional to $p=1-\phi^{*}(\mathbf{m})\phi(\mathbf{m})$. If it does not couple then a new informon is created and the likelihood that it will not couple will again be proportional to $1-x$ for some non-zero $x$. The probability that it will not couple after two plays is proportional to $p(1-x)$ and this will rapidly tend to zero with successive game play. Thus almost certainly at some point a coupling will be initiated between the particle and the apparatus. 

The wave function thus determines the likelihood of coupling between the particle and measurement apparatus processes. It will also determine the coupling to many different processes. Its interpretation as a probability distribution over causal manifold, however, is contextual, requiring the presence of a suitable measurement device whose measurement values correspond to positions in the causal manifold. The correspondence then is a secondary emergent one, requiring the intermediary measurement apparatus.

Suppose that this current particle informon couples to the current measurement apparatus informon. Then the global process undergoes a transition to an interactive process which is substantially reduced from the original process, namely

\begin{displaymath}
\tilde w_{kk'}\mathbb{P}_{k}'(\lambda_{k})\boxtimes \mathbb{M}_{k'}(\lambda_{k'})
\end{displaymath}

or in the case of no error to

\begin{displaymath}
\tilde w_{k}\mathbb{P}_{k}'(\lambda_{k})\boxtimes \mathbb{M}_{k}(\lambda_{k})
\end{displaymath}

. 

The particle process has undergone a transition which means that the descriptor no longer refers to the original generative process but only to the current process. As a consequence there will no longer be any play to those informons whose descriptors are for the previous process and its subprocesses. The same is true for the measurement apparatus as its generative process is now restricted to  the interactive version of $\mathbb{M}_{k}(\lambda_{k})$. 
Note that if the particle exits the measurement apparatus and is observed again by an identical measurement apparatus it will again couple to the same component of the second measurement apparatus but if it should be observed by a new apparatus then a new coupling will arise as a result of the transition that will be induced in the generating process by the interaction with the new measurement apparatus.

At every stage of play only a single informon is manifest, providing a determinate reality, but as these informons lie below the level of resolution of any physical apparatus only the coherent whole of the resultant tapestry and its state is actually observable. Repeated plays of the game will yield different outcomes. The apparent stochasticity of the quantum system arises from the inherent non-determinism of the game, not from an inherent indeterminism in the actual occasions themselves - the level of ultimate reality. Moreover the collapse that is observed in the wave function is not a collapse at the level of ultimate reality, which only ever manifests single informons, but rather at the level of process and merely reflects that fact that any interaction between processes potentially alters those processes.
 
The Schr\"odinger cat paradox can be understood as arising from the conflation of actual occasion and process in the Schr\" odinger formulation described previously. Prior to the release of the cyanide following the radioactive decay of the trigger material, the cat and the poison device play out uncoupled processes, and thus can be described by a free game of the form $\mathbb{P}(\text{cat})\otimes \mathbb{P}(\text{poison})$. There is no superposition of cat and device because there is no physical interaction taking place between the cat and the device. Unless the poison is released, the cat will remain alive unless it dies from natural causes waiting for the experimenter to release it. Once the poison has been released, however, the game changes and there is now an interactive game between the cat and the device of the form $\mathbb{P}(\text{cat dying})\boxtimes \mathbb{P}(\text{poison released})$. The eventual outcome of this is a dead cat. Although in NRQM it is permissible to construct a superposition state of the form $\Psi(\text{live cat})\Psi(\text{no release}) + \Psi(\text{dead cat})\Psi(\text{release})$ this is not possible in the game setting and this would need to be written as  $\Psi(\text{live cat})\Psi(\text{no release}) \oplus \Psi(\text{dead cat})\Psi(\text{release})$ reflecting the fact that there is an interaction between the two games, namely that once the second game is played it is no longer possible to play the first game. The Schr\"odinger cat paradox is thus a consequence of the limitations of the functional analytic representational system and is not a statement about reality. 

The difficulty in a paradox like Hardy's paradox arises because of an attempt to apply a wave function level statement to informon level events. Again there is no paradox, just a misapplication of a formalism to an inappropriate level. So long as the system is not disturbed the photon process will indeed generate photon informons in both arms of the apparatus but only one informon at a time, so that it appears as though the photon is both wave and particle simultaneously. But these are reflections of different levels of observation and the apparent paradox arises when the two are improperly conflated.

One last point concerns the issue of "quasi-locality". The informons being generated to form the new causal tapestry utilize information that is entirely local to each individually. The content set refers to prior informons, but this refers only to the original source of the information. It is presumed that there is a forward causal chain linking every informon within the content set to the informon and it is understood that such information has been passed forward along that causal chain to be incorporated into the informon as it comes into existence. Thus the information may have originated in the past but it reaches the informon in the present as it is formed. Therefore it is actually local to the informon at hand. Moreover, the informons that constitute any actual occasion are space-like separated, and this is true of the informons as they are formed to create the new causal tapestry. No information passes between informons within the same causal tapestry. Again, whatever information they use in their formation is local. Thus the informons form a set of local hidden variables. They are described, however, as being quasi-local because the process or processes that generate them act non locally in that they are free to add to the content of any informon on any given play, regardless of which informons have already been generated and where they are located in space-time. The processes exist in a mathematical universe that is outside of space and time. Game play may be thought of as taking place on a time scale infinitesimally smaller than the Planck time (or perhaps in a second time similar to Bars's two time physics \cite{Bars} or stochastic quantization \cite{SQ}). Note that while the processes may generate any informon on any play, no information is passed between these informons and so the non local actions of the processes do not induce any non local information exchange between informons and thus the constraints of special relativity are not violated.  

\section{A Simple Two Slit Experiment}

To illustrate the ideas above let us consider a simple model of a free non relativistic particle passing through a two slit detector and then impinging on a plate detector. This will illustrate the basic ideas of the game approach. There are two players here, one corresponding to the particle and the other to the detector. For simplicity, assume that the space is two dimensional, 0the source of the particle is at the point on the manifold denoted $(0,0)$, the two slits lie in the $x=a$ plane while the detector lies in the $x=b$ plane. The slits span the two regions from $y=c$ to $y=d$ and from $y=-c$ to $y=-d$. Again for simplicity we shall assume that the embedding to the manifold is to a regular lattice whose spatial spacing is on the order of Planck length $(l_{P})$ and temporal spacing on the order of Planck time $(t_{P})$.  

Denote the game of particle as $\mathbb{P}$. The game is played for $n$ plays. The players alternate play with the particle player playing first. The play of the detector is simple since nothing happens except to restate the presence of the detector until it is triggered by an interaction with the particle. 
In considering the play of the particle, suppose that the prior tapestry has already been generated. Particle is then free to choose any prior informon, say $[n]<\alpha>\{G\}=[n]<((t_{n},x_{n},y_{n}),\phi_{n},\mathbf{p})>\{G\}$ where $(t_{n},x_{n},y_{n})$ is the lattice embedding, $\phi_{n}$ is the partial wave function contribution, and $\mathbf{p}=\{p,\pi\}$ is a set of properties including those local to the specific actual occasion ($p$) and those that connect it to the process that generated it ($\pi$).   The partial wave function has the form $\Theta_{n}sinc(\alpha)$ where $\Theta_{n}$ is a constant and $sinc(\alpha)$ is shorthand for the product of sinc functions centered around the embedding point of $\alpha$. Let the new informon be $[m]<\beta>\{K\}$ where $\beta =((t_{n}+t_{P}, x_{n}+m_{x}l_{P},y_{m}+m_{y}l_{P}),\phi_{m},\mathbf{q})$ for integers $m_{x},m_{y}$, $\mathbf{q}=\{q,\pi\}$ and $K=\{[n]<\alpha>\{G\}\}$.

Since the particle is free after passing through the slits we can take the Lagrangian for the particle to be $L=mv^{2}/2$. Let $S[\beta,\alpha]$ denote the action calculated along a straight line path from the embedding point of $\alpha$ to the embedding point of $\beta$ over the time interval from $t_{n}$ to $t_{n}+ l_{P}$. Then the new wave function contribution has the form $\phi_{m}=e^{\frac{i}{\hslash}S[\beta,\alpha]}T_{t_{P}}T_{m_{x}l_{P}}T_{m_{y}l_{P}}\phi_{n}$.
Suppose that the embedding point has already been chosen in several previous turns. Let the informon associated with it at the current play be $[m]<((t_{n}+t_{P}, x_{n}+m_{x}l_{P},y_{m}+m_{y}l_{P}),\phi_{m})>\{G\}$ where $G$ is now a set of prior informons. Then if at the current play particle extends $[n]<\alpha>\{G\}$ to this same embedding point, then this informon is updated as follows. $G$ is replaced by $G\cup \{[n]<\alpha>\{G\}\}$. The updating of $\phi_{m}$ is rather complicated and so we illustrate it first in one dimension. The extrapolation to high dimensions follows fairly easily. In one dimension then suppose that the points of $G$ embed to the lattice points $x_{1},\ldots ,x_{k}$. Assume that these are ordered and assume, in this ordering, that $\beta$ lies between $x_{k}$ and $x_{k+1}$. Then we set $\phi_{m}= \frac{1}{A}(e^{\frac{i}{\hslash}S[\beta,t_{2}]}(t_{2}-t_{1}) + \ldots + e^{\frac{i}{\hslash}S[\beta,\alpha]}(\alpha-t_{k}) +\newline e^{\frac{i}{\hslash}S[\beta,t_{k+1}]}(t_{k+1}-\alpha) + \dots + e^{\frac{i}{\hslash}S[\beta,t_{n}]}(t_{n}-t_{n-1}))sinc(\beta)$.  $A$ is the Feynman-Hibbs normalization factor which here has the form $A=(\frac{2\pi i\hslash t_{P}}{m})^{1/2}$.

Play continues until the allotted number of allowed plays is reached or the game has already ended. Note that a player does not have to make a move on any given play if no appropriate move is available.

Note further that the contents of informons will naturally separate out into those corresponding to a passage through one slit and those corresponding to passage through the second slit. Separating out these contributions, one can naturally divide the original wave function into two distinct wave functions corresponding to these two histories and the final wave function will therefore be a superposition of these individual wave functions and the expected interference in probabilities will also occur.

This works until the particle reaches the detector plate. At that point game play by particle is able to influence game play by detector. In particular,
play by detector is actually a sum of plays, one for each location along the detector. So we should properly write the game as a sum of games so when moves by particle bring events physically close to the detector we write moves as $\sum_{x\in I}\sum_{y\in I}c_{x,y}\Pi_{x}\otimes\Delta_{y}$ where $\Pi_{x}$ are particle moves to spatial location $x$ and $\Delta_{y}$ are detector moves indicating a detection at spatial location $y$ and $c_{x,y}$ represent a form of coupling constant indicating the degree of association between moves. This coupling is expected to be a function of the wave function of the particle at $x$ and of the detector at $y$. Usually this is given as a local integral of a form $\int \Psi^{*}\mathcal{A}\Psi dz$ where $\Psi$ is the particle wave function and $\mathcal{A}$ an operator representing the action of the detector. The idea is that the detector acts upon the particle resulting in a transformation of its wave function and the product gives the probability of finding the particle therefore in such a state. Here though something different occurs. 
The position detector is arrayed upon a two dimensional surface which is parallel to the plate containing the slits and set at a fixed distance downstream. The detector possesses constant depth. The details of the detector are not relevant. It is only necessary that, however the detector is constructed, the energy of a particle interacting with the detector be dissipated through the depth of the detector in a direction roughly perpendicular to the surface. There should be minimal dissipation parallel to the surface so that when the final measurement is \textquoteleft read\textquoteright, it corresponds roughly to the location of the initial interaction. The degree of parallel spreading determines the resolution of the detector. Divide the spatial surface of the detector into (possibly overlapping) regions $\mathcal{B}_{\textbf{x}}$, each being a ball centered on the point $\textbf{x}$. Let us model the detector game $\mathbb{D}$ as a sum of games, $\sum_{\textbf{x}}\mathbb{H}_{\mathcal{B}_{\textbf{x}}}$, each game generating the informons associated with the internal activities of the detector in the spatial cylinder centered on the point $\textbf{x}$. The idea is that some stimulus takes place somewhere at the top of the cylinder and activity propagates through the cylinder to the opposing surface which registers the event, perhaps as some form of mark. In the absence of a triggering event, the play of each game is rather simple, generating a replica of the existing informon advanced in time according to the time slice of the current tapestry.

In the absence of interaction, this sum of games is rather trivial since the moves are null in all cases, which simply means that the detector is passive. Prior to any interaction, the particle and detector games are played as a normal free product, $\mathbb{P} \otimes \mathbb{D}$, meaning that moves of particle and detector are independent of one another.

Suppose now that at a particular play, particle generates, for the first time, an actual occasion that lies within the region of the detector. Let the informon so created be $[n]<(\textbf{x},\phi)>\{G\}$. The local wave function associated with this informon is $\phi$, which has the form $\Psi(\textbf{x})sinc(\textbf{z}-\textbf{x})$. Let the informons of the detector have the form $[a]<(\textbf{y},\rho)>\{K\}$ where the points $\textbf{y}$ lie within the volume occupied by the detector. Each of these informons is generated by one of the detector games, say  $\mathbb{H}_{\mathcal{B}_{\textbf{x}}}$, according to whether or not $\textbf{y}\in \mathcal{B}_{\textbf{x}}$. Then the original process is replaced by a new process $\mathbb{P}' =\mathbb{P} \otimes (\sum_{\{\mathbf{y}, \mathbf{x}\in \mathcal{B}_{\mathbf{y}}\} } \mathbb{H}_{\mathcal{B}_{\mathbf{y}}})$. This new process generates informons originating at  $[n]<(\textbf{x},\phi)>\{G\}$ while the remaining informons of the prior tapestry no longer contribute since they correspond to the original process $\mathbb{P}$ which is no longer active. The nature of the detector is such that the combined process $\mathbb{P}'$ will concentrate informons in a small region around $\mathbf{x}$ and eventually to a single detector game, say $\mathbb{H}_{\mathcal{B}_{\mathbf{y}}}$ which results in a position measurement of $\mathbf{y}$ being recorded. 

The onset of interaction between particle and detector changes the game being played. At the point of interaction the game shifts from a free product to an interactive product. The coupling between particle game and detector game is determined in part by the strength of the particle process at the point $\textbf{x}$. We assume that there is no bias in the detector so that the strength of the detector process is uniform across the detector. The coupling will depend in part upon the momentum and energy of the particle. Obviously particles that have a momentum not directed into the detector or of too high an energy will not be captured. The probability of interaction will therefore be a function of the detector and of the particle. Let the detector contribution be given by some operator $\mathcal{V}$ which acts upon the wave function to project onto possible detector states and this couples to the particle wave function as $\Psi^{*}\mathcal{V}\Psi$. The coupling probability is proportional to this term. Now this is a local coupling but because of the difference in scale between the particle and the detector we must integrate over the region of the local game. Thus the probability will be proportional to the volume integral

\begin{displaymath}
\int_{\mathcal{B}_{\textbf{y}}}\phi^{*}\mathcal{V}\phi d\nu
\end{displaymath}  

If the particle process couples to the detector process it will eventually transition to some simple coupled game of the form $\mathbb{P}\otimes \mathbb{H}_{\mathcal{B}_{\textbf{y}}}$. When this happens, the detector essentially acts like a giant slit centered on the region $\mathcal{B}_{\textbf{y}}$ and the particle process will proceed concentrated within this region. If it does not couple then play will continue according to the current combinatorial sum, which will generate another particle informon and other detector informons. Note that with repeated play the probability of the particle not coupling trends to zero and given the astronomically large number of plays this means that it will couple with certainty somewhere. Moreover the probability that this somewhere corresponds to the region $\mathcal{B}_{\textbf{y}}$ is given as above, and so corresponds to the spatial probability distribution given by the wave function according to the standard Born interpretation. However the wave function in this case determines the coupling of processes - the probability aspect is derived and reflects the particulars of the particle-detector interaction. Thus the probability in this case is emergent and contextual, as required of a non-Kolmogorov probability.

\section{Relationship to Hidden Variable Theorems}

Informons provide a form of quasi-local hidden variable. Each informon represents a discrete entity within its causal tapestry and as interpreted in the causal manifold but its interpretation in the state space is \textquoteleft fuzzy \textquoteright $\:$ and has the character of an extended object. Nevertheless the information necessary to determine a move lies internal to the informon, residing in its content. That content appears to be nonlocal but in fact the information contained therein is precisely such information as could be transmitted from the causal past to the site in the causal manifold at which the current informon manifests. That is why the information can be viewed as already being incorporated into the informon. In that sense it is local. There remains an element of nonlocality, however, in the manor of play which (borrowing a term from the author Alfred Bester) jauntes from one space-time location to another. But no information is transmitted through this nonlocality. Indeed in the space of processes there is no space-time. The processes and games exist outside of space-time as do their actions in manifesting game play. One could think of game play as taking place in a second time dimension as in  stochastic quantization \cite{SQ} or two time physics \cite{Bars}. Equally one can consider game play taking place on a time scale infinitesimally smaller than the Planck scale so that the complete play of a game takes place within an interval of duration equal to the Planck time. One should view process as being atemporal and it is the creation of actual occasions that results in an emergent space-time. The jaunting associated with process does not violate special relativity since it conveys no signal and does not occur in space-time but rather pre-space-time. 

These informons are closer in ontological status to the individual events of Bohm's model than to hidden variables per se. Nevertheless it is necessary to ensure that they do lead to violations of the various hidden variable theorems \cite{Hemmick}. Those come in two main forms - those of Bell and Leggett-Garg type involving relationships among various correlation functions based upon the wave function, and those of K-S and GHZ type, which involve relations among various measurement operators. In the reality game approach, measurements apply only to the emergent wave functions and not to the informons themselves, which manifest at scales well below those of any physical apparatus. K-S and GHZ type results assume that the functional analytic formulation holds at the lowest level of reality, but this is an assumption not a fact. If indeed quantum phenomena are emergent phenomena then there is no a priori reason to assume this. Indeed in the model presented here, the dynamics at the level of ultimate reality is described by combinatorial forcing games with tokens, and the standard functional analytic formulation is viewed as an emergent interpretation of this lower level dynamic. Indeed informons are associated with partial state functions of the form $\Psi(x)\mathrm{T}_{x}\Phi(z)$ and there is no requirement that $\Phi(z)$ be an eigenstate of any self-adjoint operator. This is clear in the case of interpolation via sinc functions. Thus the argument used in the K-S and GHZ theorems does not apply here since the measurement operators cannot be applied to these ultimate states.

Bell and Leggett-Garg type inequalities are all ultimately derived from the wave function, so it suffices to demonstrate that the state interpretation of the event tapestry approximates the wave function of the quantum system well enough to avoid violating the quantum mechanical predictions for these inequalities. However from the results presented above, we know that for Planck scale lattice spacings, the Yao and Thomas theorem suggests that the truncation error is approximately $2 \times 10^{-151}/TL^{3}$, where $T,L$ are the durations and spatial extensions of the observations. Even with errors in calculating $\Psi_{i,j}(x)$, the interpolated wave function will reproduce all of the usual quantum mechanical calculations, including the Bell and Leggett-Garg inequalities.

\section{Conclusion}

When NRQM was  initially formulated nearly a century ago, the mathematics used for its expression was a minimal extension of that used in classical mechanics. The shift was from differentiable functions on point sets to linear operators on Hilbert functions spaces. Although beautiful mathematically, the embodiment of NRQM in functional analytic language has been a mixed blessing, leading to many philosophical and technical problems. It is said that even Von Neumann himself was dissatisfied with the representation. In particular it has led some to consider notions of reality that are anthropomorphic, solipsistic and even nihilistic, denying that there even is reality. Many of these arguments derive from the (mis) application of Bell's Theorem. Following the insights of Palmer and Khrennikov, it now appears that there is a fundamental flaw in Bell's argument, namely the implicit assumption that the statistical structure of any form of \emph{classical} hidden variables must conform to the laws of Kolmogorov probability theory. This assumption is false, and depends upon the dynamics and types of interactions governing these variables. Functional analysis is limited algebraically to only single forms of addition and multiplication which leads to a conflating of event and process factors in the model. This in turn results in many paradoxes. Shifting to a particular class of combinatorial forcing games with tokens results in a non-Kolmogorov probability structure and enables one to exhibit a quasi-local, realist, hidden variable model capable of reproducing NRQM. Although the model exhibited in this paper is overly simplistic in many respects, it nevertheless demonstrates in principle that by shifting to the mathematics of combinatorial games it becomes possible to bring realism back into quantum mechanics. The model 0incorporates ideas of process theory: becoming, persisting, fading away and a transient now, all compatible with relativity. The price to be paid is to postulate the existence of an underlying layer of ultimate reality manifesting at Planck scales and which is inherently unobservable. This lower level is discrete, but fuzzy, and motion consists of discontinuous saltatory jauntes. Observable phenomena, events and space-time itself, are emergent from this lowest level. The experience of continuity is also emergent, arising though a process of interpolation. NRQM must be viewed as an effective theory, not an ultimate theory. 

Following Whitehead's process theory, the model proposes that there are two components to reality: the actual occasions which superpose to create through interpolation an emergent observable reality, and the processes that generate these occasions. These processes exist in an aspatial, atemporal algebraic universe whose structure conforms in part to a configuration of Lie algebras. The value of the combinatorial game structure is that it allows for the possibility of a non-Kolmogorovian probability structure even though the events are classical in nature. The role of the players in the game is entirely heuristic, facilitating the implementation of the dynamics without any ontological assumptions. The game setting provides a much richer variety of sums and products corresponding to the myriad ways in which games may be combined, which models more closely the diverse ways in which physical processes may combine. These additional forms of linearity and of products are necessary to resolve the paradoxes that plague NRQM.

The advantage to this model is that one obtains a richer mathematical structure, free of paradoxes, realist in nature, and conceptually simpler than current models. It has a fundamentally discrete character which may help to eliminate the infinities that hound quantum mechanics, especially quantum field theory. It provides a realist view of ultimate reality which is quasi-local (local at the level of actual occasions but non-local at the level of process or game play) and which avoids anthropomorphic and solipsistic thinking and ever more fanciful constructions that stretch credulity. In doing so it follows the principle of Occam's razor, finding a simpler model that produces the same results. The disadvanage of the model is that one must accept that quantum phenomena are emergent upon an underlying (inherently unobservable) layer of reality manifesting at Planck scales, and that quantum mechanics, like general relativity, must be understood as an effective theory. In this model continuity is the illusion and time has a definite orientation and flow. This appears to be a small price to pay for the eradication of the measurement problem and the various paradoxes, for eliminating wave-particle duality, for potentially eliminating the various infinities and for giving a sense of reality back to quantum phenomena. The next step is to attempt to extend this model into RQM and eventually to QFT. That work is hampered in part by the current state of the art in interpolation but should lead to fruitful interchanges between that branch of mathematics and quantum physics and provides abundant opportunities for future research. Perhaps in the future we will also develop better models of process which transcend the game approach.

Thanks are due to Irina Trofimova for many lengthy and fruitful discussions about emergence, process, and the fundamental role of relations and interactions, and to Robert Mann for for his sage advice and boundless patience.

\section{Appendix}

\subsection{Causal Tapestry}

The construction of a causal tapestry is recursive with the ground set assumed but usually left unspecified for simplicity. Here we shall denote the ground set by $I_{p}$ and its set of informons by $L_{p}$ and refer to it as the prior set (prior referring to the construction process, not necessarily an indicator of any temporal relationship).

Definition: Let $K$ be an index set of cardinality $\kappa$, $M$ a causal space, $I_{p}$   a union of causal tapestries. Then a (strict) causal tapestry $I$ is a 4-tuple $(L, K, M, I_{p})$ where $L$ is a set of informons (loci) satisfying the following consistency conditions:

\begin{enumerate}
\item   Each informon in $L$ has the form $[n]<\alpha>\{G\}$ with $n \in K$, $\alpha \in M$ and $G$ an acyclic directed graph whose vertex set is a subset of $L_{p}$ .
\item   The union of all $H$ such that $[n]<\alpha>\{H\}$ lies in $L$, i.e. $\cup \{H | [n]<\alpha>\{H\} \in L\}$ forms an acyclic directed graph.
\item   $[n]<\alpha>\{G\}$, $[n]<\alpha^{\prime}>\{G^{\prime}\} \in L$ implies $G = G^{\prime}$ and $\alpha = \alpha^{\prime}$
\item  $[n]<\alpha>\{G\}$, $[n^{\prime}]<\alpha>\{G\} \in L$ implies that $n = n^{\prime}$
\item  $[n]<\alpha>\{G\} \in L$ implies that $[n]<\alpha>\{G\}$ is not an element of $G$
\item   $[n]<\alpha>\{G\},[n^{\prime}]<\alpha^{\prime}>\{G^{\prime}\} \in L$ implies that neither $[n]<\alpha>\{G\} \in G^{\prime}$ nor $[n^{\prime}]<\alpha^{\prime}>\{G^{\prime}\} \in G$
\item  $[n^{\prime}]<\alpha^{\prime}>\{G^{\prime}\},[n^{\prime\prime}]<\alpha^{\prime\prime}>\{G^{\prime\prime}\} \in 0G$ and for some $[n^{\prime\prime\prime}]<\alpha^{\prime\prime\prime}>\{H\}$ there is $[n]<\alpha>\{G\} \in H$, then $[n^{\prime}]<\alpha^{\prime}>\{G^{\prime}\} \in H$ implies that $[n^{\prime\prime}]<\alpha^{\prime\prime}>\{G^{\prime\prime}\} \in H$ 
\item  The mapping $i: L_{p} \cup L \rightarrow M$ given by $i([n]<\alpha>\{G\})=\alpha$ is a causal embedding, meaning that it should be an injective causal order preserving map.
\end{enumerate}

The identifier of $[i]<\alpha>\{G\}$ is $i$, the content is $G$ and is denoted by $c(i)$ and the interpretation is $\alpha$ and is denoted by $e(i)$.

Definition: Let $K$ be an index set of cardinality $\kappa$, $\mathfrak{M} = M \times P$ a mathematical structure such that $M$ is a causal space, $P$ a generic mathematical structure termed attributes, $I_{p}$ a union of causal tapestries with attributes. Then a causal tapestry with attributes is a 4-tuple $(L, K, \mathfrak{M}, I_{p})$ such that $(L, K, M, I_{p})$ is a causal tapestry

In the causal tapestry formalism, each informon is labelled by its interpretation, and so each interpretation should carry information about the spatiotemporal associations of the informons as well as linkages to the dyadic tapestry. 
\subsection{Event and Process Tapestries}

The tapestry corresponding to the space of actual occasions, the event tapestry is defined as follows:

Def0inition: An event tapestry $\Omega$ is a causal tapestry with attributes, that is, a 4-tuple $(L, K, \mathfrak{M}, I_{p})$ where $K$ is an index set of cardinality $\kappa$, $\mathfrak{M} = M \times F(M) \times D \times P(M')$ a mathematical structure with $M$ a causal space, $F(M)$ a function (state) space, either Banach or Hilbert, $D$ a space of descriptors (properties), $P(M^{\prime})$ either a Lie algebra or tangent space on a manifold $M^{\prime}$,  $I_{p}$ a union of event tapestries. The event tapestry serves as a generalization of the notion of a position space.

Likewise we may define a process tapestry as follows:

Definition: A process tapestry $\Pi$ is a causal tapestry with attributes, that is, a 4-tuple $(L^{\prime}, K^{\prime}, \mathfrak{M}^{\prime}, I^{\prime}_{p})$ where $K^{\prime}$ is an index set of cardinality $\kappa$, $\mathfrak{M}^{\prime} = M^{\prime} \times F(M^{\prime}) \times D^{\prime} \times P^{\prime}(M)$ a mathematical structure with $M^{\prime}$ a causal space, $F(M^{\prime})$ a function (state) space, either Banach or Hilbert, $D^{\prime}$ a space of descriptors, $P^{\prime}(M)$ either a Lie algebra or tangent space on a causal manifold $M$, $I^{\prime}_{p}$ a union of process tapestries. The process tapestry serves as a generalization of the notion of a tangent space (and perhaps as a generalized momentum space as well).

Together these tapestries form an interlinked dyad.

Thus the interpretation for each informon in the event tapestry $\Omega$ takes the form $(x,f,d,p)$ where $x$ is an element of the causal manifold $M$ (spatial location), $f$ is a function on the causal manifold (state or trajectory), $d$ is a vector of descriptors such as mass, charge, and $p$ is0 an allowable transition on the process tapestry. A similar interpretation holds for each informon of the process tapestry.

An important concept in both mathematics and physics is that of duality. Duality refers to two distinct mathematical structures whose properties are nevertheless intertwined by some mathematical operation (an involution) that serves as its own inverse. The state space and momentum space representations in quantum mechanics are dual to one another and related by means of the Fourier transform. Event and process tapestries also exhibit an aspect of duality, although it is less exact except in special cases.

Let $\Omega$ be an event tapestry $(L, K, M \times F(M) \times D \times P(M^{\prime}), I_{p}  )$ and $\Pi$ a process tapestry $(L^{\prime}, K^{\prime}, M^{\prime} \times F(M^{\prime}) \times D^{\prime} \times P^{\prime}(M), I^{\prime}_{p} )$ with their structures appropriately interlinked. 

Let $C(\Omega)$ denote the graph formed by taking the union over all content sets of informons of $\Omega$ and $C(\Pi)$ denote the graph formed by taking the union over all content sets of informons of $\Pi$. These are both acyclic directed graphs by virtue of the consistency conditions for causal tapestries. 

Consider first $C(\Omega)$. Each vertex a serves as end vertex for a set of edges $E( ,a)$, and initial vertex for a set of edges $E(a, )$. Each edge in the event tapestry represents a flow of information from the initial to the terminal vertex arising through the process of origination of the informon as a consequence of the play of the reality game. The process that enables this flow of information to take place is one that manifests in reality, and thus corresponds to a physically realizable transition or transformation. This transformation is represented in the archetype interpreting the causal tapestry as either a global symmetry operation or as a local tangent. This in turn should correspond to an informon in the process tapestry. Thus we expect that there should be a mapping of the edge set of $C(\Omega)$ into the vertex set of $C(\Pi)$. Likewise, each edge in the process tapestry marks a flow from one transition to another transition and each such transition between transitions requires some form of manifestation of an actual occurrence to express it. This reflects the fact that the presence of an actual occurrence has a meaningful impact upon reality, and that impact manifests itself as an influence shaping the particular manner in which the next actual occasion will come into being. That process of coming into being, being marked by some transition, will therefore be altered by the manifesting of an actual occurrence, and so some new process may become manifest. Thus it makes sense that each transition between transitions should correspond to some informon which manifests the information that influences the appearance of the next transition. Thus it is reasonable to expect that there should exist a mapping from the edge set of $C(\Pi)$ into the vertex set of $C(\Omega)$. 

Unfortunately it will usually be the case that each vertex of either graph will have more than one edge for which it is terminal or initial. In order to get around this problem let us define a new graph called a covering graph. Let $G = {V,E}$ be a generic directed graph.
For each vertex $v \in V$, define a new vertex set by $\{e( ,v)\} \times \{e(v, )\}$. Let $V(CG(G))$ consist of the union of all such vertex sets for $v \in V$ together with all terminal vertices of $V$ (a vertex is terminal if it is not the initial vertex of any edge). For any pair of such vertices in $V(CG(G))$, $e(a,v)e(v,b)$ and $e(c,v^{\prime})e(v^{\prime},d)$, define 00an edge $e(e(a,v)e(v,b),e(c,v^{\prime})e(v^{\prime},d))$ if and only if $b=c$. For any terminal vertex $v \in V$, define an edge $(v,\phi)$. Define the new edge set $E(CG(V))$ to be the union of all of the above edges taken over all of $V$.

Definition (Tableau Duality Postulate): Let $\Omega$ and $\Pi$ be event and process tapestries respectively. Let $C(\Omega)$ and $C(\Pi)$ denote their respective content graphs. Then there exist subgraphs $C^{\prime}(\Omega)$ and $C^{\prime}(\Pi)$ of $CG(C(\Omega))$ and $CG(C(\Pi))$ respectively such that there exists an order isomorphism between $C^{\prime}(\Omega)$ and $C^{\prime}(\Pi)$ with $V(C^{\prime}(\Omega))$ mapping to $E(C^{\prime}(\Pi))$ and $E(C^{\prime}(\Pi))$ mapping to $V(C^{\prime}(\Omega))$.

Moreover, in general we require an additional criterion to be met. 

Definition (Tableau Consistency Criterion): Let $\Omega$ and $\Pi$ be event and process tapestries respectively which satisfy the tableau duality postulate. Let $[n]<\alpha>\{G\}$ and $[n^{\prime}]<\alpha^{\prime}>\{G^{\prime}\}$ be informons within the content set of some informon of $\Omega$ and assume that $[n]<\alpha>\{G\} \rightarrow [n^{\prime}]<\alpha^{\prime}>\{G^{\prime}\}$ in the content graph. By the tableau duality assumption there will exist an informon $[i]<\beta>\{H\}$ within the content set of some informons of $\Omega$ such that this edge maps to this informon. Now $\alpha = (m,f,d,g)$ and $\alpha^{\prime} = (m^{\prime},f^{\prime},d^{\prime},g^{\prime})$ and $\beta =(s,e,h,l)$. Then we also require that $m^{\prime}=l(m)$. A similar result holds for $\Pi$. This ensures that if each informon is interpreted as an element of the causal manifold, then each edge is interpreted as a transformation from one element to the next. 

Definition: A reality tableau consists of an event tapestry $\Omega$ and a process tapestry $\Pi$ that satisfy the tableau duality postulate together with the tableau consistency criterion.

The processes involved in the creation of actual occasions will now generate a succession of reality tableaux. This occurs through the repeated play of a reality game. 

\section{The Reality Game}

As in the case of forcing games, the reality game requires the completion of a set of tasks. 

Consider an event tapestry $\Omega$ first as the case for $\Pi$ is analogous. Assume that the index set $K$ has sufficient cardinality to account for all possible informons, past, present and future. In a universe of discrete events, a countable $K$ will usually suffice. Otherwise it can generally be taken to be $2^{\kappa}$ where $\kappa$ is the cardinality of the informon set of the current tapestry.

First of all, a new set of informons must be generated. Each informon has the form $[n]<\alpha>\{G\}$ with $n \in K$, $\alpha \in \mathfrak{M}$ and $G$ an acyclic directed graph whose vertex set is a subset of $L \cup L_{p}$ . 

First, select an unused element $n^{\prime} \in K$ and form a bare informon $[n^{\prime}]<>\{\}$. Next, a content must be selected. The elements of the new 0content set must lie in $L \cup L_{p}$.

In general, in order to ensure that the consistency conditions are properly met, each content set will consist of the union of an up-set of a prior content set together with one or more informons from the current tapestry included as maximal elements. An up-set $\hat G$ of an ordered set $G$ is a subset of $G$ such that if $x$ lies in $\hat G$ then every element $y$ in $G$ 
such that $x<y$ also lies in $\hat G$. The new content set may have only one current informon, which means a simple extension. It may have several current informons, which means an interaction is about to take place. It may also be the case that several new informons may possess the exact same content (though of course the labels and interpretations will differ), which corresponds to the case in which multiple processes originate from a given informon (such as occurs following an interaction).

If $G$ is an ordered set, let $\hat G$ denote an up-set of $G$. If $G$ and $H$ are ordered sets, define the ordered sum $G + H$ to be the ordered set such that every element of $H$ is greater than every element of $G$. Then the different possible outcomes take the form:

\begin{enumerate}
\item Given an informon $a = [n]<\alpha>\{G\}$, add an informon of the form $[n^{\prime}]<\alpha^{\prime}>\{\hat G + \{a\}\}$
\item Given a collection of informons $\{a_{i} = [n_{i}]<\alpha_{i}>\{G_{i}\}\}$, add an informon of the form $[n^{\prime}]<\alpha^{\prime}>\{\cup_{i}\hat G_{i} + \{a_{i}\}\}$
\item Given an informon $[n]<\alpha>\{G\}$, add a collection of informons $\{a_{i} = [n_{i}]<\alpha_{i}>\{G_{i}\}\}$ where $G_{i} = \hat G + \{a\}$ (n00ote $\hat G$ may be different for different $G_{i}$ )
\end{enumerate}

Each informon of $\Omega$ must also be interpreted within its mathematical structures. These interpretations must meet additional criteria:

\begin{enumerate}
\item  $[n]<\alpha>\{G\}$ and $[n]<\alpha^{\prime}>\{G\}$ in $L$ implies $\alpha=\alpha^{\prime}$
\item  $[n]<\alpha>\{G\}$ and $[n^{\prime}]<\alpha>\{G\}$ in $L$ implies that $n = n'$
\item   Let $[n]<\alpha>\{G\}$ and $[n^{\prime}]<\alpha^{\prime}>\{G^{\prime}\}$ be informons within the content set of some informon of $\Omega$ and assume that $[n]<\alpha>\{G\} \rightarrow [n^{\prime}]<\alpha^{\prime}>\{G^{\prime}\}$ in the content graph. There exists an informon $[i]<\sigma>\{H\}$ within the content set of some informon of $\Pi$ such that if $\alpha = (m,f,d,g)$ and $\alpha^{\prime} = (m^{\prime},f^{\prime},d^{\prime},g^{\prime})$ and $\sigma =(s,e,h,l)$, then $m^{\prime}=l(m)$.
\item  The mapping $i: L_{p} \cup L \rightarrow M$ given by $i([n]<\alpha>\{G\}) = i((x,f,d,p)) = x$ is a causal embedding, meaning that it should be an injective causal order preserving map.
\item  Let the mapping $m: L_{p} \cup L \rightarrow F(M)$ be given by $m([n]<\alpha>\{G\}) = m((x,f,d,p)) = f$. Define an interpretation of the content $G$, denoted $m(G)$ by setting $m(G) = \cup_{G} m(a)$ $(a \in G)$. 
\item  The game dynamics relative to any particular informon should depend upon its interpretation, the interpretation of its content (which renders its meaning in terms of the archetypal interpretation), as well as the interpretations of surrounding informons. The game dynamic may involve additional constraints such as conservation laws, symmetry rules, extremal principles as so on.
\end{enumerate}

Conditions 4 and 5 provide the critical criteria which any reality game must ensure are satisfied. Interpolation theory provides the means to satisfy them. Interpolation provides the essential bridge between the discrete structure the informons of the reality tableaux and the continuous structure of their interpretations in the causal manifold and in the state (or trajectory) space. The correct choice of the embedding into the causal manifold which determines the sampling, and the frame functions (such as $sinc(x)$ above) which generate the interpolation into the function space, will depend upon the details of each individual model. Only sinc interpolation will be discussed here but there are more general approaches to interpolation for a wide variety of function spaces, including an approach specifically for graphs such as causal tapestries \cite{jorgensen}. Generally speaking, the sampling need only be uniformly distributed throughout the space-like hypersurface into which the informons embed. The frame functions will depend upon the symmetries and boundary conditions of the problem. The sampled values can be generated through an application of Green's functions techniques (which are determined again by symmetries and boundary conditions) to extend from prior informons to these new informons. In this way the reality game enables the entire history of a system to be generated, or bootstrapped from a set of initial conditions. 

A useful heuristic tool for working with reality games is the token, which is the formal analogy of a game piece such as a chess piece, or checker piece. The generation of an actual occasion may occur in several steps, particularly when non-local contributions from multiple prior occasions must be integrated 00into a new occasion. The use of tokens permits the reality game to keep track of this information until game play is completed, at which point the tokens may be "exchanged" for the completed actual occasion. Tokens constitute Whitehead\textquoteright s "unobservable parts" making up the "indecomposable whole" of an actual occasion.

Assume that the set of descriptors, $D$, for some causal tapestry possesses some algebraic structure and that the elements of $D$ can be generated from some subset, $B$, using those algebraic operations. A set of tokens, $T$, is a set of elements together with a mapping $T \rightarrow B$ which assigns values to each token. A reality game with tokens is a reality game together with a set of tokens, $T$, where at the end of each move, some token $t$ will be attached to some bare informon $[n ]<>\{\}t_{1}\cdots…t_{n}\rightarrow [n ]<>\{\}t_{1}\cdots…t_{n}t$, which might possibly have some sequence of tokens $t_{1}\cdots…t_{n}$  already attached. At the end of play this token sequence will be stripped off and an interpretation and content inserted.

The dynamics of a physical process is implemented by means of a co-operative, two player, combinatorial, forcing game with tokens, the \emph{reality game}, which generates a succession of reality tableaux $(\Omega,\Pi),(\Omega_{1},\Pi_{1}),(\Omega_{2},\Pi_{2}), \ldots$, unfolding the history of a system. Dynamical laws are formulated as strategies for the players of this game. Games are used to represent the properties and actions of processes. The role of the \emph{player} is entirely heuristic, enabling the implementation of the game dynamic, representing the intrinsic, implicit and necessary functionality required of process for the creation of informons. The concept of the player is not meant to be taken literally and there is no implication as to the existence of any form of intentionality or agency. 

Play proceeds for a fixed number of steps prescribed in advance during which players alternate in performing one of a set of tasks on either tapestry of the reality tableau. Tasks for Player 1 include choosing an index (thus creating a bare informon $[k]<>\{\}$), constructing the content $G$ of an informon, and laying down tokens, which serve both construction purposes and denote various properties. The content $G$ can have one of three forms depicted in the following graphs, where k and l represent newly created informons and informons m and n come from $L \cup L_{p}$. Here,(a) corresponds to single process extension, (b) to process inactivation, and (c) to process activation.

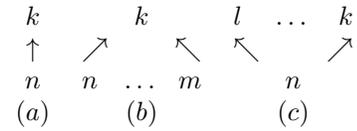
\begin{displaymath}
\begin{array}{cclcrclcr}
k &   &   & k &   &   & l & \dots & k \\
\uparrow &  & \nearrow &  & \nwarrow &  & \nwarrow &  & \nearrow \\
n  &   & n & \dots & m &   &   & n &  \\
(a) & & &(b) &  & & & (c) &
\end{array}
\end{displaymath}

Tasks for Player 2 include choosing an element of the causal manifold, choosing a weight for the interpolation basis element and choosing a transformation element. For both players, each move must satisfy various consistency criteria. At a given step a player may choose not to move. The game is a win for Player 2 if the global state obtained by summing the states of all informons in the tapestry interpolates the correct wave function to within some predetermined tolerance. The goal is to identify strategies for Players 1 and 2 which ensure that Player 2 always wins. 

Strategies of interest for Player 2 build interpretations following an idea of Kempf \cite{Kempf} for the construction of continuity from discreteness based on Shannon sampling theory. In its simplest version, Player 2 embeds informons into the causal manifold $M$ so as to form a sampling of $M$, and embeds them into the state space $\mathcal{H}(M)$ using a biorthogonal basis composed of translates of a template function $\Phi$. Each informon contributes a term $\Psi(x_{i})\mathrm{T}_{x_{i}}\Phi$ to a global wave function obtained by summing contributions over the informons (with or without their contents). The usual representation of a state as a linear superposition $\Psi = \sum_{i} w_{i}\Psi_{i}$ of eigenfunctions of some operator $A$ is not useful here as the weights $w_{i}$ are defined as integrals over the entire causal manifold $M$ whereas a reality tableaux corresponds only to a discrete sampling of a space-like hypersurface of $M$. However, the weights $\Psi(x_{i})$ can be defined for individual informons and calculated using variants of Green's functions or path integral methods applied to the current tapestry (including content sets in time dependent settings).

%Create the reference section using BibTeX:

\end{document}